\makeatletter
\@ifundefined{@parse@version@dash}{%
\def\@parse@version#1{\@parse@version@0#1}
\def\@parse@version@#1/#2/#3#4#5\@nil{%
\@parse@version@dash#1-#2-#3#4\@nil}
\def\@parse@version@dash#1-#2-#3#4#5\@nil{%
  \if\relax#2\relax\else#1\fi#2#3#4 }
}{} 
\makeatother
 
\documentclass[%
 reprint,
superscriptaddress, 
nofootinbib,
 amsmath,amssymb,    
 aps,
 ]{revtex4-2}

\usepackage{color}
\usepackage{graphicx}
\usepackage{dcolumn}
\usepackage{bm}
\RequirePackage{siunitx}   		
\DeclareSIUnit\eVperc{\eV\per\clight}
\DeclareSIUnit\clight{\text{\ensuremath{c}}}
\RequirePackage{booktabs}  
\RequirePackage{multirow}		
\RequirePackage{pgf}
\RequirePackage{hyperref}

\newcommand{\cotan}[1][]{\textnormal{cotan}{\,#1}}
\newcommand{\ii}{\textrm{i}}
\newcommand{\ti}{\textrm{\tiny i}}
\newcommand{\intdif}[3]{\int_{#1}^{#2}\!\!\!\textrm{d}{#3}}

\newcommand{\expo}[1]{\textrm{e}^{#1}}

\newcommand{\diff}[3][]{\frac{\partial^{#1}{#2}}{\partial{#3}^{#1}}}

\newcommand{\nuc}[2]{{}^{#1}{\textnormal{#2}}}

\newcommand{\spc}[3]{{}^{#1}{#2}_{#3}}

\newcommand{\braket}[2]{\left\langle{#1}\right|\left.{#2}\right\rangle}
\newcommand{\couple}[3]{\left[{#1}\!\times\!{#2}\right]^{#3}}

\newcommand{\ket}[1]{\left|{#1}\right\rangle}

\newcommand{\mc}[1]{\multicolumn{1}{c}{{#1}}}

\begin{document} 


 \title{A study of the fine-structure constant dependence of radiative capture in Halo-EFT}

\author{Ulf-G.~Mei{\ss}ner}
\email{meissner@hiskp.uni-bonn.de}%
\affiliation{Helmholtz-Institut~f\"{u}r~Strahlen-~und~Kernphysik,%
~Rheinische~Friedrich-Wilhelms~Universit\"{a}t~Bonn,~D-53115~Bonn,~Germany}
\affiliation{Bethe~Center~for~Theoretical~Physics,%
~Rheinische~Friedrich-Wilhelms~Universit\"{a}t~Bonn,~D-53115~Bonn,~Germany}
\affiliation{Institute~for~Advanced~Simulation~(IAS-4),%
~Forschungszentrum~J\"{u}lich,~D-52425~J\"{u}lich,~Germany}
%
\author{Bernard~Ch.~Metsch}%
\email{metsch@hiskp.uni-bonn.de}%
\affiliation{Institute~for~Advanced~Simulation~(IAS-4),%
~Forschungszentrum~J\"{u}lich,~D-52425~J\"{u}lich,~Germany}
\affiliation{Helmholtz-Institut~f\"{u}r~Strahlen-~und~Kernphysik,%
~Rheinische~Friedrich-Wilhelms~Universit\"{a}t~Bonn,~D-53115~Bonn,~Germany}
\author{Helen~Meyer}%
\email{hmeyer@hiskp.uni-bonn.de}%
\affiliation{Helmholtz-Institut~f\"{u}r~Strahlen-~und~Kernphysik,%
~Rheinische~Friedrich-Wilhelms~Universit\"{a}t~Bonn,~D-53115~Bonn,~Germany}
\affiliation{Bethe~Center~for~Theoretical~Physics,%
~Rheinische~Friedrich-Wilhelms~Universit\"{a}t~Bonn,~D-53115~Bonn,~Germany}

\date{\today}

\begin{abstract}
We study the fine-structure constant dependence of the rates of some
selected radiative capture reactions within the framework of so-called
Halo Effective Field Theory in order to assess the adequacy of some
assumptions made on the Coulomb penetrability.  We find that this
dependence deviates from that implied by a parameterization of the
cross sections of this effect via a simple penetration factor. Some
features of this fine-structure dependence are discussed, in
particular its potential impact on the abundances of the light
elements in primordial nucleosynthesis.
\end{abstract}

\maketitle


\section{\label{sec:intro}Introduction}
In Ref.~\cite{Meissner:2023voo} we made a re-assessment of the
electromagnetic fine-structure constant dependence of the light element abundances 
in primordial nucleosynthesis or Big Bang nucleosynthesis
(BBN). This required a description of the fine-structure constant
dependence of the pertinent cross sections of the leading reactions in
the BBN network. Only for the leading nuclear reaction, \textit{i.e.}
the radiative capture reaction $p + n \to d + \gamma$ a detailed and
sufficiently accurate theoretical description within the framework of
pionless Effective Field Theory (EFT) is available,
see~\cite{Rupak:1999rk}. For the other reactions we relied on a
parameterization of the fine-structure constant dependence that
accounted for the dependence of $Q$-values of the nuclear reactions
through changes in the nuclear binding energies due to the Coulomb
interaction of the protons as well as a modeling of the Coulomb
penetration factors in the form
\begin{equation}
  \label{eq:penetration}  
  P(x) = \frac{x}{\expo{x}-1}
\end{equation}
with
\begin{equation}
  \label{eq:eta}
  x
  =
  2\pi\,\frac{Z_a\,Z_b\,\mu_{ab} c^2\,\alpha}{c\,p}
  =
  \sqrt{\frac{E_G(\alpha)}{E}}
\end{equation}
in terms of the so-called Gamow energy for a two-particle reaction channel $ij$
\begin{equation}
  \label{eq:GamowE}
  E_G(\alpha)
  =
  2\,\pi^2\,Z_i^2\,Z_j^2\,\mu_{ij}^{} c^2\,\alpha^2
\end{equation}
and the center-of-mass (CMS) energy $E$ or $E+Q$ for the entrance and
the exit channel, respectively. Here, $p$ is the corresponding CMS
momentum, $Z_i$ the charge (in units of the elementary charge $e$) of
nuclide $i$, $\mu_{ij}$ the reduced mass, $c$ is the speed of light and $\alpha$ denotes the
fine-structure constant.  In addition we accounted for a simple linear
dependence on $\alpha$ in case of radiative capture reactions as well
as a trivial $\alpha$ dependence reflecting the final momentum
dependence if assuming dominance of dipole radiation,
see~\cite{Meissner:2023voo}. We also noted in~\cite{Meissner:2023voo}
that for some other radiative capture reactions an effective field
theory description, \textit{viz.}  ``Halo-EFT'', is available that
potentially offers the possibility to study the $\alpha$ dependence of
the cross sections analytically and thus assess the validity of the
assumptions made in~\cite{Meissner:2023voo}.
For a comprehensive overview on applications of Halo-EFT to nuclear
structure and reactions we refer to the review~\cite{Hammer:2017tjm}.
The purpose of the present paper is to
study the $\alpha$ dependence of the cross sections and the
corresponding rates for the following radiative capture reactions: The
neutron induced reaction
\begin{equation}
  \label{eq:reacn7Li8Lig}
  n + \nuc{7}{Li} \to \nuc{8}{Li} + \gamma\,,
\end{equation}
as treated in Refs.~\cite{Fernando:2011ts,Higa:2020kfs}, 
the proton induced reaction
\begin{equation}
  \label{eq:reacp7Be8Bg}
  p + \nuc{7}{Be} \to \nuc{8}{B} + \gamma\,,
\end{equation}
as treated in Ref.~\cite{Higa:2022mlt} and the two reactions that are
most relevant to BBN:
\begin{equation}
  \label{eq:reac3H4He7Lig}
  \nuc{3}{H} + \nuc{4}{He} \to \nuc{7}{Li} + \gamma\,,
\end{equation}
and 
\begin{equation}
  \label{eq:reac3He4He7Beg}
  \nuc{3}{He} + \nuc{4}{He} \to \nuc{7}{Be} + \gamma\,,
\end{equation}
as treated in Refs.~\cite{Higa:2016igc,Premarathna:2019tup}.
Although in the present contribution we shall rely on the inplementation as
elaborated in Refs.~\cite{Fernando:2011ts}-\cite{Premarathna:2019tup}
we like to mention that the
$n + \nuc{7}{Li} \to \nuc{8}{Li} + \gamma$
reaction was also treated within the Halo-EFT framework 
in Ref.~\cite{Zhang:2014zsa}. Furthermore, earlier Halo-EFT work on the
$p + \nuc{7}{Be} \to \nuc{8}{B} + \gamma$ reaction can be found in
Refs.~\cite{Zhang:2015ajn,Zhang:2017yqc}. For a discussion on the
$\nuc{3}{He} + \nuc{4}{He} \to \nuc{7}{Be} + \gamma$ reaction,
we refer to~\cite{Zhang:2019odg}. 

The paper is organized as follows: In Sect.~\ref{sec:bf} we
recapitulate the formulas for the radiative capture cross section in
Halo-EFT. We then compare the results for the nominal $\alpha$ value
with experimental data in Sect.~\ref{sec:cs}. The results on the
$\alpha$ dependence of the cross sections or astrophysical $S$-factors
and the corresponding rates are discussed in Sect.~\ref{sec:alphadep}.
The impact on the changes of the light element abundances with a
variation of the fine-structure constant is presented in
Sect.~\ref{sec:abun}.  We summarize our findings in Sect.~\ref{sec:summary}.  Some
technicalities not given in
Refs.~\cite{Higa:2022mlt}-\cite{Premarathna:2019tup} are relegated to
the Appendices.

\section{\label{sec:bf}Basic formalism}

In Halo-EFT the nuclear system is assumed  to consist of a
``core''-system with mass $m_c$\,, charge number $Z_c$ and spin $s_c$
and a ``valence''-system with mass $m_v$\,, charge number $Z_v$ and
spin $s_v$\,. Furthermore, $M=m_c+m_v$ and $\mu=m_c\,m_v / M$ denote
the total mass  and the reduced mass of the system, respectively.  With $L_i$
denoting the orbital angular momentum of the relative motion in the
initial state the total spin and angular momentum in the initial state
is given by $\vec{S}_i = \vec{s}_c+\vec{s_v}$ and ${\vec J}_i =
\vec{S}_i + \vec{L}_i$\,, respectively. Within the effective range
expansion the initial state interaction is then specified by the
parameters $a_{\zeta_i}$, $r_{\zeta_i}$, and $s_{\zeta_i}$, ($\zeta_i
= \spc{S_i}{L_i}{J_i}$) which for $L_i=0$ correspond to the $s$-wave
scattering length, effective range and the first shape parameter,
respectively.

The final state in the radiative capture reaction has mass $M_f$\,,
charge number $Z_f$, excitation energy $E_x$ and total nuclear spin
$J_f$\,.  In terms of partial waves $\spc{S_f}{L_f}{J_f}$ the final state is
written as
\begin{eqnarray}
  \label{eq:finalstate}
  \ket{M_f,E_x;J_f}
  &=&
  \sum_{s_f,L_f}
      a_{S_f,L_f}\,\ket{\couple{\couple{s_c}{s_v}{S_f}}{L_f}{J_f}}
      \nonumber\\
  &=&
  \sum_{\zeta}
  a_{\zeta_f}\,\ket{\zeta_f}
\end{eqnarray}
with $S_f$ the total spin of the di-nuclear-cluster and $L_f$ the
relative orbital angular momentum quantum number.
The coefficients $a_{S_f,L_f}$ are the amplitudes for the decomposition
of the final state in terms of di-nuclear states. Then
$$B_\zeta=M-(M^\zeta_f+E_x^\zeta)$$ is called the separation energy
with respect to the clusters ``$v$'' and ``$c$'' and
$$\gamma_\zeta=\sqrt{2\,M\,B_\zeta}$$ the binding momentum of this
state. We define
\begin{equation}
  \label{eq:kc} k_C = Z_c\,Z_v\,\alpha\,\mu
\end{equation}
as the inverse Bohr radius of a di-nuclear system in the case of
charged particles.

\subsection{\label{sec:sigmaE1}Electric dipole radiative capture}

The formulas in this section are adopted from
Ref.~\cite{Premarathna:2019tup}.  Assuming that the radiative capture
proceeds through an electric dipole transition and that only a single
state contributes, the cross section is given by the expression
\begin{eqnarray}
  \label{eq:csE1}
  \lefteqn{
  \sigma_{E1}(p)
  =
  \frac{1}{16\pi\,M^2}\,
  \frac{1}{(2s_c+1)(2s_v+1)}
  }
  \nonumber\\
&&\hspace*{6em}
   \times
   \sum_\zeta |a_\zeta|^2\,
   \frac{k^{(\zeta)}_\gamma}{p}\,\left|M_{E1}^{(\zeta)}\right|^2\,,
\end{eqnarray}
where $p$ is the magnitude of the relative momentum in the CMS with
$E=p^2/(2\mu)$ the non-relativistic expression for the energy of
the relative motion and
\begin{equation}
  \label{eq:gammamom}
  k^{(\zeta)}_\gamma = \frac{p^2+\gamma_\zeta^2}{2\,\mu}
\end{equation}
the non-relativistic approximation to the momentum of the photon in
the final state.

The dimensionless amplitude squared reads
\begin{eqnarray}
  \label{eq:amplitudesp}
  \left|M_{E1}^{(\zeta)}\right|^2
  &=&
      64\pi\,\alpha\,(2J^\zeta_f+1)\,
      \frac{
      \left(
      Z_v m_c
      -
      Z_c m_v
      \right)^2}{\mu\,\gamma_\zeta}
      \mathcal{N}(\eta^\zeta_\gamma,\rho^\zeta_\gamma)
      \nonumber\\
  &&
  \times\left[
    \left|\mathcal{A}(p)\right|^2
    +
    2\,\left|Y(p)\right|^2
  \right]\,.
\end{eqnarray}
Here, we defined $\eta^\zeta_\gamma = k_C/\gamma_\zeta$ and
$\rho^\zeta_\gamma = \rho_1^\zeta/\gamma_\zeta$ with $\rho_1^\zeta =
\hbar c/r_1^\zeta$ the effective momentum and $r_1^\zeta$ the
effective range in the channel $\zeta$\,. The normalisation is given
by
\begin{eqnarray}
  \label{eq:norm}
  \lefteqn{
  \mathcal{N}(\eta,\rho)
  }
  \nonumber\\
  &=&
      \frac{2\pi}{%
      -
      \rho
      +4\,\eta\,h(\eta)
      +2\,\eta^2\,(\eta^2-1)\,h'(\eta)
      }\,,
\end{eqnarray}
where
\begin{equation}
  \label{eq:hfunc}
  h(\eta) = \psi(\eta) + \frac{1}{2\,\eta}- \log(\eta)
\end{equation}
and $\psi(\eta) = \Gamma'(\eta)/\Gamma(\eta)$ is the digamma
function.
The normalisation is thus completely determined by the binding energy
of the di-nuclear cluster (via $\eta$) and the effective momentum (via
$\rho$) in the final state.
In case of a neutral cluster $k_C = 0$ and thus $\eta =
k_C/\gamma = 0$\,. Then
\begin{equation}
  \label{eq:norm0}
  \left.\mathcal{N}(\eta,\rho)\right|_{\eta=0}
  =
  -\frac{2\pi\,\gamma}{\rho+3\,\gamma}\,.
\end{equation}

With $\eta_p=k_C/p$ the capture from the initial $s$-wave is
given by the amplitude
\begin{equation}
  \label{eq:amplitudeAs}
  \frac{\left|\mathcal{A}(p)\right|}{C_0(\eta_p)}
  =
  \left|
    X(p)
    -
    \frac{2\pi}{\mu^2}\,
    \frac{B(p) + \mu\,J_0(p)+\mu^2\,k^{(\zeta)}_\gamma\,L^{(\zeta)}_{E_1}}{
      \left[C_0(\eta_p)\right]^2\,p\left(\cotan{(\delta_0)}-\ii\right)}
  \right|\,,
\end{equation}
where $L^{(\zeta)}_{E_1}$ is the low-energy constant of the the two-body current contact term and 
\begin{eqnarray}
  \label{eq:amplitudeX}
  X(p)
  &=&
      1+ \frac{2}{3}\,\kappa\,\frac{\Gamma(2+\eta_\gamma)}{C_0(\eta_p)}\,
      \intdif{0}{\infty}{\rho}\,\,
      W_{-\eta_\gamma,\frac{3}{2}}\left(2\,\kappa\,\rho\right)\,
      \nonumber\\
  &&\hspace*{3em}\times
      \left[
      -\frac{F_0(\eta_p,\rho)}{\rho}
      +
      \partial_\rho F_0(\eta_p,\rho)
      \right]\,,
\end{eqnarray}
is the $s$-wave contribution without initial state strong interactions
in terms of Coulomb functions $F_\ell$ and Whittaker functions
$W_{\eta,\mu}$, which are the solutions to the pure Coulomb problem.
In the case where $k_C=0$ (\textit{i.e.} if a neutral particle is
involved) this reduces to
\begin{equation}
  \label{eq:amplitudeXp}
  \left.X(p)\right|_{\eta_\gamma=0}
  =
  1 - \frac{2}{3}\,\frac{p^2}{p^2+\gamma^2}\,.
\end{equation}
The $s$-wave contribution from strong initial-state interactions is given by
\begin{eqnarray}
  \label{eq:BJ}
  \frac{B(p) + \mu\,J_0(p)}{\mu^2\,p}
  &=&
      \frac{1}{3\pi}\,\frac{\ii-\kappa^3}{1+\kappa^2}
      +
      \eta_p\,\frac{C(p)}{\mu^2} + \frac{\Delta B(p)}{\mu^2\,p}
      \nonumber\\
  &&
      \!\!-
      \frac{\eta_p}{2\pi}
      \left[
      2\,h(\ii\,\eta_p) + 2\,\gamma_E - \frac{5}{3}+\log{(4\pi)}
     \right]\,.
     \nonumber\\
\end{eqnarray}
Here, the function $C(p)$ is given by a double integral, treated in
App.~\ref{sec:CInt} and the finite contribution ${\Delta B}(p)$ is
evaluated as follows: The integrand
\begin{eqnarray}
  \label{eq:BIntgndp}
  \mathcal{B}(\kappa,\eta_\gamma;\rho)
  &=&
      -\frac{\kappa}{3\pi}
      \Gamma(2+\eta_\gamma)\,\Gamma(1+\ii\,\kappa\,\eta_\gamma)\,
      W_{-\eta_\gamma,\frac{3}{2}}\left(2\,\kappa\,\rho\right)\,
      \nonumber\\
  &&
     \hspace*{-5em}
     \times\,\left[
     -\frac{1}{\rho}\,W_{-\ti\kappa\eta_\gamma,\frac{1}{2}}(-2\,\ii\,\rho)
     +
     \partial_\rho W_{-\ti\kappa\eta_\gamma,\frac{1}{2}}(-2\,\ii\,\rho)
     \right]
\end{eqnarray}
where $\kappa = \eta_p/\eta_\gamma=\gamma/p$\,, in 
\begin{equation}
  \label{eq:Bintgrl}
  \frac{B(p)}{\mu^2\,p}
  =
  \intdif{0}{\infty}{\rho}\,\,\mathcal{B}(\kappa,\eta_\gamma;\rho)\,,
\end{equation}
is quadratically divergent for $\rho \to 0$\,. Noting that the
integrand depends on $\alpha$ via $\eta_\gamma = k_C/\gamma$
and $k_C \propto \alpha$\,, the integrand can be
regularized by subtracting the terms from zero and single
photon contributions, \textit{i.e.}  the terms of
$\mathcal{O}(\alpha^0)$ and $\mathcal{O}(\alpha^1)$. Then,
with
\begin{displaymath}
  \alpha\,\diff{\mathcal{B}(\alpha)}{\alpha}
  =
  \eta_\gamma\,\diff{\mathcal{B}(\kappa,\eta_\gamma;\rho)}{\eta_\gamma}\,,
\end{displaymath}
the finite contribution\footnote{
  As pointed out in Ref.~\cite{Premarathna:2019tup} the divergent
  pieces of $B$ cancel the divergent pieces in $J_0$, this was accounted
  for in arriving at Eq.~(\ref{eq:BJ})\,.
}
is given by
\begin{eqnarray}
  \label{eq:DeltaB}
  \frac{{\Delta B}(p)}{\mu^2\,p}
  =
  \intdif{0}{\infty}{\rho}\,\,
  &&
  \Bigl[
  \mathcal{B}(\kappa,\eta_\gamma;\rho)
     \nonumber\\
  &&
     \hspace*{-3em}
     -
    \mathcal{B}(\kappa,0;\rho)
    -
    \left(\partial_{\eta_\gamma}\mathcal{B}\right)(\kappa,0;\rho)\cdot\eta_\gamma
  \Bigr]
\end{eqnarray}
which can be integrated numerically\footnote{%
  The partial derivative involved might be difficult to find
  analytically. Although in principle not particularly stable, one could
  use numerical approximations to the (partial) derivative of a function
  $\mathcal{B}(\kappa,\eta_\gamma;\rho)$\,, such as (with $\mathcal{B}_j
  = \mathcal{B}(\kappa,j\,h;\rho)$)\,:
  \begin{eqnarray*}
    \left.(\partial_{\eta_\gamma}
    \mathcal{B}(\kappa,\eta_\gamma,\rho)
    \right|_{\eta_\gamma=0}
    &=&
        \frac{8\,(\mathcal{B}_{+1}
        -
        \mathcal{B}_{-1})-(\mathcal{B}_{+2}-\mathcal{B}_{-2}))}{12\,h}
        +
        \mathcal{O}(h^4)
    \\
    &&\hspace*{-9em}
    =
       \frac{45\,(\mathcal{B}_{+1}-\mathcal{B}_{-1})
       -
       9\,(\mathcal{B}_{+2}-\mathcal{B}_{-2})
       +
       (\mathcal{B}_{+3} - B_{-3}) }{60\,h}
       +
       \mathcal{O}(h^6)
  \end{eqnarray*}
  for some small finite $h$\,.
}.
For neutral particles, \textit{i.e.} $k_C=0$, one obtains
\begin{equation}
  \label{eq:BJp}
  \left.\frac{B(p)+\mu\,J_0(p)}{\mu^2\,p}\right|_{k_C=0}
  =
  \frac{1}{3\pi}\,\frac{\ii-\kappa^3}{1+\kappa^2}-\frac{\ii}{2\pi}\,.
\end{equation}

Finally, the contribution from the initial $d$-wave states to the 
capture process is given by the amplitude
\begin{eqnarray}
  \label{eq:amplitudeY}
  Y(p)
  &=&
      \frac{2}{3}\,\kappa\,\Gamma(2+\eta_\gamma)\,
      \intdif{0}{\infty}{\rho}\,\,
      W_{-\eta_\gamma,\frac{3}{2}}\left(2\,\kappa\,\rho\right)\,
      \nonumber\\
  &&\hspace*{3em}\times
      \left[
      \frac{2\,F_2(\eta_p,\rho)}{\rho}
      +
      \partial_\rho F_2(\eta_p,\rho)
      \right]
\end{eqnarray}
which for $k_C=0$ reduces to
\begin{equation}
  \label{eq:amplitudeYp}
  \left.Y(p)\right|_{k_C=0}
  =
  \frac{2}{3}\,\frac{p^2}{p^2+\gamma^2}\,.
\end{equation}

\subsection{\label{sec:M1}Magnetic dipole radiative capture}

In case of single nucleon radiative capture there are additional
relevant contributions from magnetic dipole transitions to the final
states.

\subsubsection{\label{sec:M1n}Neutron induced magnetic dipole contribution}

In case of the $n + \nuc{7}{Li} \to \nuc{8}{Li} + \gamma$ reaction
we recapitulate the formulas from  Ref.~\cite{Higa:2020kfs}.
Earlier work on this reaction, including the $M1$-contribution,
can be found in Ref.~\cite{Fernando:2011ts}.

The cross section for the $M1$ contribution to the radiative capture
in the $\nuc{7}{Li} + n \to \nuc{8}{Li} + \gamma$ reaction through the
$3^+$ resonance according to Ref.~\cite{Fernando:2011ts} is given by
\begin{eqnarray}
  \label{eq:sigmaM1Li}
  \sigma_{M1}(p)
  &=&
      \frac{1}{14}\,\frac{7}{3}\,
      \frac{\alpha\,\mu}{m_p^2}\,
      \left|h^2\,\mathcal{Z}^{(\zeta)}\right|\,
      \left(\frac{k}{p}\right)^3\,
  \nonumber\\
  &&\hspace*{-3em}
     \times
     \left|
     \frac{p^2}{-\frac{1}{a_1^{(3)}+\frac{1}{2}\,r_1^{(3)}\,p^2}-\ii\,p^3}
     \right|^2
     \Biggl\lbrace
     \left|
     \frac{2}{3}\,\frac{\gamma^3-\ii\,p^3}{\gamma^2+p^2}\,K^{(2)} + \beta^{(2)}
     \right|^2
      \nonumber\\
  &&\hspace*{5em}
     +
     \left|
     \frac{2}{3}\,\frac{\gamma^3-\ii\,p^3}{\gamma^2+p^2}\,K^{(1)} + \beta^{(1)}
     \right|^2
     \Biggr\rbrace\,,
\end{eqnarray}
with $m_p$ the proton mass. The asymptotic normalization of the
final $\nuc{8}{Li}$-states ($\zeta=2^+$ for the ground state or
$\zeta=1^+$ for the first excited state) with binding momentum
$\gamma^{(\zeta)}$ is given by 
\begin{displaymath}
  h^2\,\mathcal{Z}^{(\zeta)} = -\frac{2\pi}{3\,\gamma^{(\zeta)}+r_1^{(\zeta)}}\,,
\end{displaymath}
the gyro-magnetic factors in the  
$\spc{5}{P}{3} \to \spc{3}{P}{2}$
and the 
$\spc{5}{P}{3} \to \spc{5}{P}{2}$
$M1$-transitions
are given by
\begin{eqnarray}
  \label{eq:M1K}
  K^{(1)}
  &=&
      \sqrt{\frac{3}{2}}
      \left(
      \frac{3}{2}\,g_c - \frac{3}{2}\,g_v
      \right)\,,
      \nonumber\\
  K^{(2)}
  &=&
      \sqrt{\frac{3}{2}}
      \left(
      \frac{3}{2}\,g_c - \frac{3}{2}\,g_v
      \right)\,,
      \nonumber\\
  K^{(2)}
  &=&
      \sqrt{\frac{3}{2}}
      \left(
      \frac{3}{2}\,g_c + \frac{1}{2}\,g_v
      +
      2\,\mu\,m_n\,\frac{Z_c}{m_c^2}
      \right)\,,
\end{eqnarray}
in terms of the gyro-magnetic ratios $g_c$ and $g_v$ describing the
magnetic moments of the core and the valence system, respectively,
and $\beta^{(i)}, i=1,2$ are constants reflecting the two-body current
terms. Finally, $a_1^{(3)}$ and $r_1^{(3)}$ are the scattering volume
and the effective momentum in the $\spc{5}{P}{3}$ scattering channel.

\subsubsection{\label{sec:M1p}Proton induced magnetic dipole contribution}

This section summarizes the results quoted in
Ref.~\cite{Higa:2022mlt} for the $M1$ contribution in the reaction
\begin{displaymath}
  \nuc{7}{Be} + p \to \nuc{8}{B}(2^+) + \gamma\,,
\end{displaymath}
through the $1^+$ resonance. Considered is only the $\spc{5}{P}{1} \to
\spc{5}{P}{2}$ transition assuming that the $1^+$ resonance
is dominantly a proton $\spc{}{p}{\frac{1}{2}}$ coupled to the
$\nuc{7}{Be}(\frac{3}{2}^-)$ ground state with the amplitude
\begin{equation}
  \label{eq:7Bestate}
  \braket{\left[
      \frac{3}{2}^-\times
      \left[\frac{1}{2}^+\times
        1^-\right]^\frac{1}{2}\right]^1}{
    \left[
      \left[
        \frac{3}{2}^-\times\frac{1}{2}^+\right]^2
      \times1^-\right]^1}
  =
  \sqrt{\frac{5}{6}}\,.
\end{equation}

The cross section for a magnetic dipole radiative reaction through the
$1^+$ resonance is then given  by~\cite{Higa:2022mlt}
\begin{equation}
  \label{eq:sigmaM1}
  \sigma_{M1}(p)
  =
  \frac{1}{16\pi\,M^2}\, 
  \frac{1}{6}
  \sum_\zeta \frac{(k_\gamma^{(\zeta)})^3}{p^3}\,\left|M^{(\zeta)}_{M1}\right|^2\,.
\end{equation}
The squared matrix element reads
\begin{eqnarray}
  \label{eq:M1M2}
  \left|M^{(\zeta)}_{M1}\right|^2
  &=&
  (2J^{(\zeta)}_f+1)\,
  \mu^3\,
  \frac{8\pi\,M^2\,\alpha}{m_p^2}\,
  \frac{|\mathcal{A}_1(p)|^2}{|C_1(\eta_p)|^2}\,
      \nonumber\\
  &&\hspace*{2em}
     \times
  \frac{2\pi}{\mu}\,\mathcal{Z}^{(\spc{5}{P}{2})}\,
  \frac{\mu^2}{432\,\pi^4}\,
  \left|\overline{L}_{22}(p)\right|^2\,,
\end{eqnarray}
where 
\begin{eqnarray}
  \label{eq:L22}
  \overline{L}_{22}(p)
  &=&
      \frac{2\pi}{\mu}
      \Biggl\lbrace
      \frac{9\pi}{\sqrt{40}}
      \Biggl[
      3\,g_c + g_v
     + 4\,\mu\,m_p
     \left(
     \frac{Z_\phi}{m_\phi^2} + \frac{Z_\psi}{m_\psi^2}
     \right)
     \Biggr]
      \nonumber\\
  &&\hspace{10em}
     \times
     \frac{\overline{D}(p,\gamma_\zeta)}{\mu^2}
     -\beta_{22}
     \Biggr\rbrace
\end{eqnarray}
with $\vec \mu_c = g_c\,\vec{j}_c$ and $\mu_v = g_v\,\vec{j}_v$ the (spin)
magnetic moments of the constituents $c\,,v$ of the di-nuclear
system and
\begin{displaymath}
  \vec \mu_L = \mu\,m_p\left( \frac{Z_c}{m_c^2} + \frac{Z_v}{m_v^2}\right)\,\vec L
\end{displaymath}
the magnetic moment due to the current associated with the relative orbital motion.
\footnote{
  Note that the current due to the velocity of fragment $i$ is given
  by the operator $(Z_i\,e)/(m_i)\,\widehat{\vec
    p}_i$\,. Accordingly the associated orbital magnetic moment, expressed
  in units of $\mu_N = (e\,\hbar c)/(2\,m_p c^2)$\,, reads $\mu_i =
  Z_i\,(m_p/m_i)\,\mu_n\,\widehat{\vec L}_i$ with $L_i$ the angular
  momentum of fragment $i$\,.
}

Furthermore
\begin{equation}
  \label{eq:Dbardl}
  \frac{\overline{D}(p)}{\mu^2\,\gamma}
  =
  \frac{1}{3\pi}\,\frac{\ii\,\kappa^3-1}{1+\kappa^2}
  +\eta_\gamma\,\frac{D'(\eta_p,\kappa)}{\mu^2}
  +\frac{\Delta D(k_C;p,\gamma)}{\mu^2\,\gamma}\,,
\end{equation}
where we defined  
$\rho = p\,r\,, \eta_\gamma = {k_C}/{\gamma}\,,
\kappa={\gamma}/{p}\,, \eta_p = \kappa\,\eta_\gamma = {k_C}/{p}$.
The integrand
\begin{eqnarray}
  \label{eq:calD}
  \mathcal{D}(\kappa,\eta_\gamma;\rho)
  &=&
  -\ii\,\frac{\kappa}{3\pi}\,
      \Gamma(2+\ii\,\kappa\,\eta_\gamma)\,\Gamma(2+\eta_\gamma)
      \nonumber\\
  &&\hspace*{0.5em}
     \times
  W_{-\eta_\gamma,\frac{3}{2}}(2\,\kappa\,\rho)\,
  W_{-\ii\,\eta_p,\frac{3}{2}}(-2\,\ii\,\rho)
\end{eqnarray}
in
\begin{equation}
  \label{eq:Dintegral}
  D(k_C;p,\gamma) = \mu^2\,p\,\intdif{0}{\infty}{\rho}\,\mathcal{D}(\kappa,n_\gamma;\rho) 
\end{equation}
is again divergent for $\rho \to 0$\,. By subtracting the zero and
single photon contributions one then defines the finite term
\begin{eqnarray}
  \label{eq:DeltaD}
  \Delta D(k_C;p,\gamma)
  &=&
  \mu^2\,p\,
  \intdif{0}{\infty}{\rho}
  \Bigl[
  \mathcal{D}(\kappa,\eta_\gamma;\rho)
  \nonumber\\
  &&\hspace*{-3em}
    -
    \mathcal{D}(\kappa,0;\rho)
    -
    \left(\partial_{\eta_\gamma}\mathcal{D}\right)(\kappa,0;\rho)\cdot\eta_\gamma
  \Bigr]~.
\end{eqnarray}
The evaluation of the second term $D'(\eta_p,\kappa)$ on the r.h.s. of
Eq.~(\ref{eq:Dbardl}) is given in App.~\ref{sec:DInt}.

The initial-state interaction in the $\spc{5}{P}{1}$ channel is given
by the amplitude
\begin{eqnarray}
  \label{eq:A1ampdl}
  \lefteqn{
  \gamma^2\,\mathcal{A}_1(p)
  =
  \frac{2\pi\,\gamma}{\mu}
  }
  \nonumber\\
&&\hspace*{-2em}
   \times
  \frac{9\,(C_1(\eta_p))^2\,\expo{\ti\,2\,\sigma_1}}{
    -\frac{1}{a_1^{(\spc{5}{P}{1})}\,p^2\,\gamma} + \frac{1}{2}\,\frac{r^{(\spc{5}{P}{1})}}{\gamma}\,
    -2\,\frac{p}{\gamma}\,\eta_p\,(\eta_p^2+1)\,H(\eta_p)}\,,
\end{eqnarray}
where $a^{(\spc{5}{P}{1})}$ is the scattering volume 
and $r^{(\spc{5}{P}{1})}$ the effective momentum to reproduce the $1^+$ resonance 
with position $E_R=p_R^2/(2\mu) = 0.630(3) \textnormal{ MeV}$ and
width $\Gamma_R = 0.0357(6) \textnormal{ MeV}$\,.
The normalization of the final state is determined by 
\begin{eqnarray}
  \label{eq:Z5P2dl}
  \lefteqn{
  \gamma\,\frac{2\pi}{\mu}\mathcal{Z}^{(\spc{5}{P}{2})}
  =
  }
  \nonumber\\
  &&\hspace*{1em}
  \frac{2\pi}{-\rho_\gamma
    + 2\,\eta_\gamma^2(\eta_\gamma^2-1)\,h'(\eta_\gamma)
      +4\,\eta_\gamma\,h(\eta_\gamma)}
\end{eqnarray}
with $\rho_\gamma = {r_1^{(\spc{5}{P}{2})}}/{\gamma}$\,.

\section{\label{sec:cs}Cross sections, astrophysical S-factors and reaction rates.}

The cross sections $\sigma$ or the corresponding astrophysical $S$-factors given as
\begin{equation}
  \label{eq:Sfac}
  S(E) = E\,\sigma(E)\,\expo{\sqrt{E_G/E}}
\end{equation}
with $E_G$ 
the Gamow energy in the entrance channel, were calculated according to
the formulas given in the previous section, Sect.~\ref{sec:bf}.
The nuclear structure parameters are given in Tab.~\ref{tab:nucpar},
while the reaction parameters for the nucleon induced reactions are
given in Tab.~\ref{tab:reacpar} and for the radiative capture to
$\nuc{7}{Li}$ and $\nuc{7}{Be}$ are given in Tab.~\ref{tab:reacpar34}.

\begin{table}[!htb]
  \centering
  \caption{Nuclear structure data: Nuclear mass in MeV,
    spin/parity $J^\pi$, excitation energy $E_x$ in MeV,
    binding momentum $\gamma$ with respect to the di-nuclear system in MeV,
    the gyro-magnetic ratio $g$ and the ground state nuclear Coulomb energy $V_C$  in
    MeV~\cite{Meissner:2023voo}.}
  \label{tab:nucpar}
  \begin{ruledtabular}
    \begin{tabular}{crrrrrr}
      Nucleus & Mass      & $J^\pi$         & $E_x$ & $\gamma$ & $g$ & $V_C$ \\
      \colrule
      $p$ &  938.2721           & $\frac{1}{2}^+$ & 0      & \mc{-}  &  5.5857 & 0 \\
      $n$ &  939.5654           & $\frac{1}{2}^+$ & 0      & \mc{-}  & -3.8261 & 0 \\
      $\nuc{3}{H}$  &           & $\frac{1}{2}^+$ & 0      & \mc{-}  & \mc{-} & 0 \\
      $\nuc{3}{He}$ & 2808.3916 & $\frac{1}{2}^+$ & 0      & \mc{-}  & \mc{-}  & 0.6884 \\
      $\nuc{4}{He}$ & 3727.3794 & $0^+$           & 0      & \mc{-}  & \mc{-}  & 0.7588 \\
      $\nuc{7}{Li}$ & 6533.8330 & $\frac{3}{2}^-$ & 0      & 88.9099 &  2.1710 & 1.5994 \\
              &           & $\frac{1}{2}^-$ & 0.4776 & 79.8429 & \mc{-}  & \mc{-} \\
      $\nuc{7}{Be}$ & 6534.1841 & $\frac{3}{2}^-$ & 0      & 71.2970 & -0.9329 & 2.7117 \\
                    &           & $\frac{1}{2}^-$ & 0.4291 & 60.8999 & \mc{-}  & \mc{-} \\
      $\nuc{8}{Li}$ & 7471.3658 & $2^+$           & 0      & 57.7872 & \mc{-} & 1.6491 \\
                    &           & $1^+$           & 0.9808 & 41.5694 & \mc{-} & \mc{-} \\
      $\nuc{8}{B}$  & 7472.3201 & $2^+$           & 0      & 14.9465 & \mc{-} & 4.2119 \\
    \end{tabular}
  \end{ruledtabular}
\end{table}

\begin{table}[!htb]
  \centering
  \caption{
    Reaction parameters for the nucleon induced radiative capture reactions:
    $s$-wave scattering length $a_0$ in fm, $p$-wave scattering volume $a_1$ in fm$^3$\,,
    $p$-wave effective momentum $r_1$ in MeV, two-body current
    parameter $\beta^{(1)}=\beta^{(2)} = \beta$ for the neutron induced
    reaction, $\beta=\beta_{22}$ for the proton induced reaction, both in
    MeV. $r_1(\spc{5}{P}{2})^*$ is the value used in calculating the $M1$
    contribution.
    For the $\nuc{7}{Li}(n,\gamma)\nuc{8}{Li}$ reaction the parameter
    sets ``A'' and ``ANC'' correspond to those called ``EFT A'' and ``EFT
    ANC'' in Ref.~\cite{Higa:2020kfs}, respectively.  For the
    $\nuc{7}{Be}(p,\gamma)\nuc{8}{B}$ reaction the parameter sets ``NNLO''
    and ``ANC'' correspond to those called
    ``$\textnormal{EFT}_{\textnormal{\tiny gs}}$ I NNLO'' and those
    related to a determination from A(symptotic) N(ormalization)
    C(oefficients), respectively, see Ref.~\cite{Higa:2022mlt}.
  }
  \label{tab:reacpar}
  \begin{ruledtabular}
    \begin{tabular}{lrrrr}
      Parameter & \multicolumn{2}{c}{$\nuc{7}{Li}(n,\gamma)\nuc{8}{Li}$} 
      & \multicolumn{2}{c}{$\nuc{7}{Be}(p,\gamma)\nuc{8}{B}$} \\
      \colrule 
      & A & ANC & NNLO & ANC \\ 
      \cline{2-3}\cline{4-5} 
      $a_0(\spc{3}{S}{1})$ &    0.87    &    0.87   & 17.34  & 17.34  \\
      $a_0(\spc{5}{S}{2})$ &   -3.63    &   -3.63   & -3.18  & -3.18  \\
      $r_1(\spc{3}{P}{2})$ & -290.0707  & -605.6995 & -173.0 & -176.8 \\
      $r_1(\spc{5}{P}{2})$ & -290.0707  & -270.0028 & -32.92 & -40.31 \\
      $r_1(\spc{5}{P}{2})^*$& -290.1    & -290.1    & -30.00 & -30.00 \\
      $r_1(\spc{3}{P}{1})$ & -473.5848  & -638.1095 & &\\
      $a_1(\spc{5}{P}{1})$ &            &           & -108.13 & -108.13\\
      $r_1(\spc{5}{P}{1})$ & -473.5848  & -498.0909 & -111.23 & -111.23\\
      $a_1(\spc{5}{P}{3})$ &  -77.0136  & -547.1    & & \\
      $r_1(\spc{5}{P}{3})$ &  -77.0136  & -547.1    & & \\
      $\beta$              & 170.0      & 170.0     &   375.0 &   375.0 \\
    \end{tabular}
  \end{ruledtabular}
\end{table}

\begin{table}[!htb]
  \centering
  \caption{Reaction parameters for the
    $\nuc{3}{H} + \nuc{4}{He} \to \nuc{7}{Li} + \gamma$ and
    $\nuc{3}{He} + \nuc{4}{He} \to \nuc{7}{Be} + \gamma$ reactions:
    The $s$-wave scattering length $a_0$ in fm,
    the $s$-wave effective range $r_0$ in fm,
    the $s$-wave shape parameter $s_0$ in fm$^3$,
    the $p$-wave effective momentum $r_1$ in MeV,
    the $p$-wave shape parameter $s_1$ in fm and the LEC.
      The parameter sets labeled `fit'', ``A'' and ``fit'',``AII''
      correspond to those labeled ``$\chi^2$'',``Model A'' and
      ``$\chi^2$'' and ``Model AII'' in Ref.~\cite{Premarathna:2019tup}
      for these two reactions, respectively.
  }
  \label{tab:reacpar34}
  \begin{ruledtabular}
    \begin{tabular}{lrrrr}
      Parameter & \multicolumn{2}{c}{$\nuc{4}{He}(\nuc{3}{H},\gamma)\nuc{7}{Li}$} 
      & \multicolumn{2}{c}{$\nuc{4}{He}(\nuc{3}{He},\gamma)\nuc{7}{Be}$} \\
      \colrule
                & fit & A & fit & AII \\ 
      \cline{2-3}\cline{4-5} 
      $a_0(\spc{2}{S}{\frac{1}{2}})$ &  17.0 & 13.0 & 22.0 & 40.0  \\
      $r_0(\spc{2}{S}{\frac{1}{2}})$ &   0.6 & -0.1 &  1.2 &  1.09 \\
      $s_0(\spc{2}{S}{\frac{1}{2}})$ &   2.0 & 11.0 & -0.9 &  -2.2 \\
      $r_1(\spc{2}{P}{\frac{1}{2}})$ & -129.0 & -230.0 & -41.9 & -45.0 \\
      $r_1(\spc{2}{P}{\frac{3}{2}})$ & -149.0 & -190.0 & -55.4 & -59.0  \\
      $s_1(\spc{2}{P}{\frac{1}{2}})$ & \mc{-} & \mc{-} & 1.74 &  1.84\\
      $s_1(\spc{2}{P}{\frac{3}{2}})$ & \mc{-} & \mc{-} &  1.59 & 1.69 \\
      LEC$(\spc{2}{P}{\frac{1}{2}})$ & 1.5  & 4.0 & 0.83 & 1.07 \\
      LEC$(\spc{2}{P}{\frac{3}{2}})$ & 1.44 & 2.2 & 0.78 & 1.02 \\
    \end{tabular}
  \end{ruledtabular}
\end{table}

The reaction rate in thermal equilibrium at temperature $T$ then
follows from the cross section via
\begin{equation}
  \label{eq:rate}
  \gamma(T)
  =
  N_A\sqrt{\frac{8}{\pi\,\mu\,(kT)^3}}\intdif{0}{\infty}{E}\,\,\,\sigma(E)\,E\,\expo{-\frac{E}{kT}}\,,
\end{equation}
where $N_A$ is the Avogadro number and $k$ the Boltzmann constant.
Essentially, this expression for the rate has the form of a
Laplace-transform of the cross section multiplied by the CM-energy.

In the next subsections we present the resulting cross sections or
astrophysical $S$-factors as well as the corresponding rates according
to Eq.~(\ref{eq:rate}) for the four reactions studied here, all
calculated at the present value of the fine-structure constant that we
shall call the nominal value of $\alpha$ given by
\begin{equation} 
  \label{eq:alpha0}
  \alpha_0 = 7.2973525693(11) 10^{-3} = 1/137.035999084(2)  
\end{equation}
from Ref.~\cite{Workman:2022ynf}.
  For each of the reactions considered here we present the calculated rates for
  two parameter sets used in the Halo-EFT calculations in order to give an impression
  of the systematic uncertainty. In addition we compare the resulting rates
  with those used in the original versions of some publicly available BBN codes: \textit{viz.}
  \texttt{NUC123}~\cite{Kawano:1992ua},
  \texttt{AlterBBN}~\cite{Arbey:2011nf,Arbey:2018zfh},
  \texttt{PArthENoPE}~\cite{Pisanti:2007hk,Consiglio:2017pot,Gariazzo:2021iiu} and
  \texttt{PRIMAT}~\cite{Pitrou:2018cgg}, 
  if available. These were also considered in our study~\cite{Meissner:2023voo}
  mentioned in the introduction.
  The most recent code \texttt{PRyMordial},
  see~\cite{Burns:2023sgx,PRyMordial-Code}, by default uses the
  \texttt{PRIMAT}-rates and thus in this context is not discussed
  separately.

\subsection{The $n + \nuc{7}{Li} \to \nuc{8}{Li} + \gamma$ reaction}

In this case $Z_v = 0, Z_c=3$\,, $s_v = 1/2\,, s_c={3}/{2}$ and
$Z_f=4$ with $J_f = 2$ for the ground state and $J_f=1$ for the
excited state.  The final $2^+$ state is supposed to be an equal
mixture of the $\spc{3}{P}{2}$ and $\spc{5}{P}{2}$ states,
\textit{i.e.}
\begin{equation}
  \label{eq:2+}
  \ket{2^+}
  =
  \frac{1}{\sqrt{2}}\ket{\spc{3}{P}{2}}
  +
  \frac{1}{\sqrt{2}}\ket{\spc{5}{P}{2}}~,
\end{equation}
while the excited $1^+$ state is supposed to be
\begin{equation}
  \label{eq:1+}
  \ket{1^+} = -\frac{1}{\sqrt{6}}\ket{\spc{3}{P}{1}} + \sqrt{\frac{5}{6}}\ket{\spc{5}{P}{1}}~.
\end{equation}

The total radiative capture cross section  in this case is given by the
sum of the expression for the electric dipole contribution given in
Eqs.~(\ref{eq:csE1},\ref{eq:amplitudesp}) with the special formulas
for $k_C=0$ given in
Eqs.~(\ref{eq:norm0},\ref{eq:amplitudeXp},\ref{eq:BJp},\ref{eq:amplitudeYp})
and the magnetic dipole contribution of Eq.~(\ref{eq:sigmaM1Li})\,.

\begin{figure}[!htb]
  \centering
  \includegraphics[width=0.475\textwidth,trim= 145 430 100 130, clip]{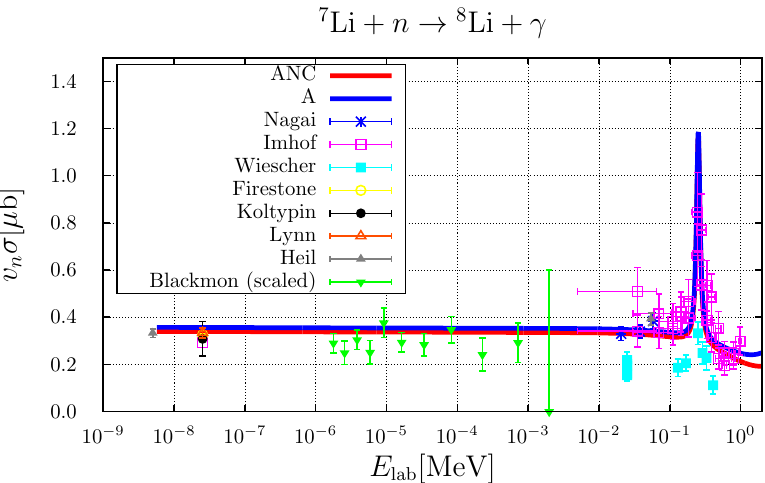} 
  \caption{
    Cross section (scaled with the laboratory neutron velocity
    $v_n/c$) as a function of the laboratory neutron energy compared to
    experimental data:
    Nagai~\cite{Nagai:2005qu},
    Imhof~\cite{Imhof:1959zz},
    Wiescher~\cite{Gorres:1989zz},
    Firestone~\cite{Firestone:2016gos},
    Koltypin~\cite{Koltypin:1956xyz},
    Lynn~\cite{Lynn:1991zz},
    Heil~\cite{Heil:1998xyz},
    Blackmon~\cite{Blackmon:1996zz}.
    The latter data were 
    divided by the branching ratio 0.89
    for the ground state in order to also
    account for the $1^+$ final state contribution
    such that these data now represent the total capture cross
    section.
  }
  \label{fig:n7Li8Lig_vnsigma}
\end{figure}

The resulting cross section (scaled with the laboratory neutron
velocity) is compared to the experimental data in
Fig.~\ref{fig:n7Li8Lig_vnsigma}.

\begin{figure}[!htb]
  \centering
  \includegraphics[width=0.475\textwidth,trim= 140 430 90 130, clip]{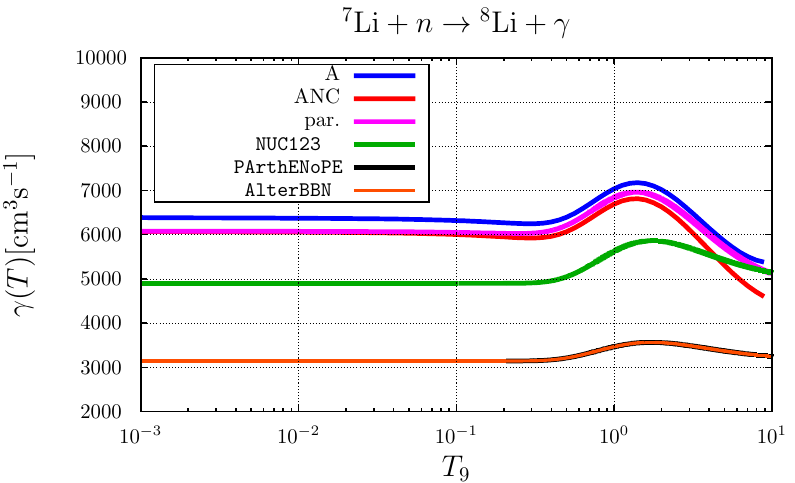}  
  \caption{
    Comparison of the temperature-dependent rate for the
    $n+\nuc{7}{Li} \to \nuc{8}{Li} + \gamma$ reaction.
    Here, $T_9 = T / 10^9~\textnormal{K}$. The blue and red
    curves represent the results from the parameter sets ``A'' and ``ANC'',
    respectively, the purple curve is based on the parameterization
    of Eq.~(\ref{eq:Spar7Ling8Li}), the green curve is the rate used in
    \texttt{NUC123}~\cite{Kawano:1992ua},
    the curves for the parameterizations in
    \texttt{PArthENoPE}~\cite{Pisanti:2007hk,Consiglio:2017pot,Gariazzo:2021iiu} (black curve)
    and \texttt{AlterBBN}~\cite{Arbey:2011nf,Arbey:2018zfh} (brown curve) are identical.
  }
  \label{fig:n7Li8Lig_rate}
\end{figure}

The calculated rates for the parameter sets ``A'' and ``ANC'' (these
correspond to the parameter sets called ``EFT A'' and ``EFT ANC'' in
Ref.~\cite{Higa:2020kfs}, respectively; ``ANC'' standing for:
``parameters corresponding to empirical A(symptotic) N(ormalization)
C(oefficient'')) are compared to the rates as parameterized in
\texttt{NUC123}~\cite{Kawano:1992ua},
\texttt{PArthENoPE}~\cite{Pisanti:2007hk,Consiglio:2017pot,Gariazzo:2021iiu}
and \texttt{AlterBBN}~\cite{Arbey:2011nf,Arbey:2018zfh}
as well as to the rate resulting from the following novel
parameterization of the cross section, accounting for the $1^+$
resonance via a non-relativistic Breit-Wigner parameterization
\begin{eqnarray}
  \label{eq:Spar7Ling8Li}
     \sqrt{E}\,\sigma(E)
  &=&
      0.0675\,\frac{1-0.045\,E+0.7\,E^2}{1+0.001\,E+0.7\,E^2}
      \nonumber\\
  &&
     +\frac{0.018}{1+5000.0\,(E-0.2215)^2}\,,
  \nonumber\\
  &&
  \textnormal{(in mb MeV${}^{\frac{1}{2}}$, with $E$ in MeV)}
\end{eqnarray}
in Fig.~\ref{fig:n7Li8Lig_rate}. Indeed this parameterization yields a rate
very similar to those of the Halo-EFT calculation.

\subsection{The $p + \nuc{7}{Be} \to \nuc{8}{B} + \gamma$ reaction}

In this case $Z_v = 1, Z_c=4$\,, $s_v = 1/2\,, s_c={3}/{2}$ and
$Z_f=5$ with $J_f = 2$ for the ground state, which, as the
corresponding ground state of the mirror nucleus is supposed to be an
equal mixture of the $\spc{3}{P}{2}$ and $\spc{5}{P}{2}$ states.  The
total radiative capture cross section is given by the sum of the
expression for the electric dipole contribution in
Eqs.~(\ref{eq:csE1},\ref{eq:amplitudesp}) with $k_C \ne 0$ and the
resonant magnetic dipole contribution as given in
Eq.~(\ref{eq:sigmaM1})\,.

\begin{figure}[!htb]
  \centering
  \includegraphics[width=0.475\textwidth,trim= 140 430 100 130, clip]{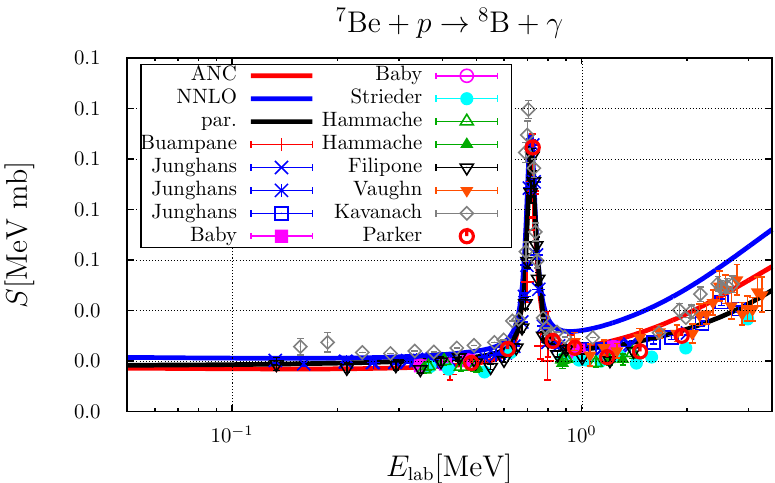} 
  \caption{$S$-factor as a function of the laboratory proton energy
    compared to experimental data:
    Buompane~\cite{Buompane:2022qbt},
    Junghans~\cite{Junghans:2010zz,Junghans:2003bd,Junghans:2001ee,Junghans:2004spv},
    Baby~\cite{ISOLDE:2002bco,ISOLDE:2002qgw,ISOLDE:2003dpg},
    Strieder~\cite{Strieder:2001ozz},
    Hammache~\cite{Hammache:1997rz},
    Filipone~\cite{Filippone:1983tkv,Filippone:1983zz},
    Vaughn~\cite{Vaughn:1970gv},
    Kavanach~\cite{Kavanach:1960xyz},
    Parker~\cite{Parker:1966zz}.
    The blue and red curve correspond to the parameter sets ``NNLO''
    and ``ANC'', respectively.  The purple curve represents the
    parameterization given in Eq.~(\ref{eq:p7Be8Bg_par})\,.
    }
  \label{fig:p7Be8Bg_S}
\end{figure}

The resulting $S$-factor 
\begin{eqnarray}
  \label{eq:p7Be8Bg_par}
     S(E)
     &=&
     0.018\,\frac{1+0.3\,E+0.125\,E^2}{1+0.017\,E^2}
  \nonumber\\
     &&\hspace*{5em}
        +\frac{0.090}{1+2500.0\,(E-0.63)^2}\,,
        \nonumber\\
  &&
  \textnormal{(in MeV mb, with $E$ in MeV)}\,.
\end{eqnarray}
is compared to experimental data and to the
parameterization of Eq.~(\ref{eq:p7Be8Bg_par}) thereof in
Fig.~\ref{fig:p7Be8Bg_S}.

\begin{figure}[!htb]
  \centering
  \includegraphics[width=0.475\textwidth,trim= 140 430 100 130, clip]{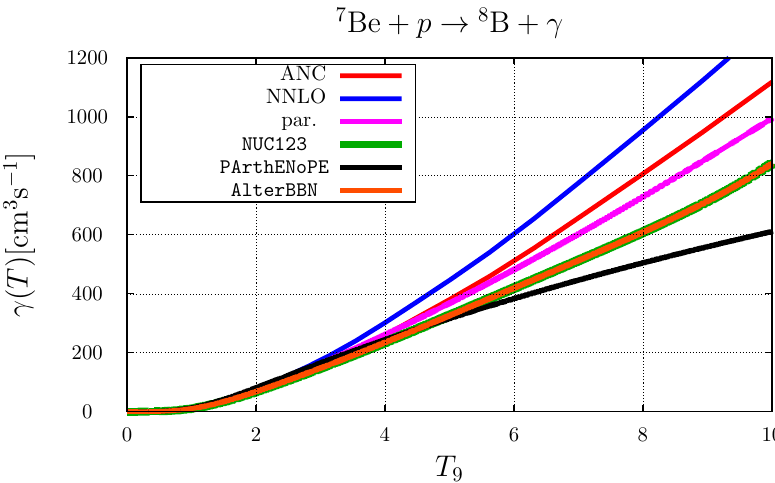}  
  \caption{
    Comparison of the temperature ($T_9 = T / 10^9~\textnormal{K}$) dependent rate  
    for the $p+\nuc{7}{Be} \to \nuc{8}{B} + \gamma$ reaction. The blue
    and red curves represent the parameter sets ``NNLO'' and ``ANC'',
    respectively, the purple curve is based on the parameterization of
    Eq.~(\ref{eq:p7Be8Bg_par}), the black curve is the rate used in
    \texttt{PArthENoPE}~\cite{Pisanti:2007hk,Consiglio:2017pot,Gariazzo:2021iiu},
    the curves for the parameterizations in
    \texttt{AlterBBN}~\cite{Arbey:2011nf,Arbey:2018zfh} (brown curve)
    and \texttt{NUC123}~\cite{Kawano:1992ua} (green curve) are identical. 
  } 
  \label{fig:p7Be8Bg_rate}
\end{figure}

The calculated rate for the parameter sets ``NNLO'' and ``ANC'', where
these labels refer to the parameter sets labeled
``$\textnormal{EFT}_{\textnormal{\tiny gs}}$ I NNLO'' and those
related to a determination from A(symptotic) N(ormalization)
C(oefficients), respectively, see Ref.~\cite{Higa:2022mlt},
are compared to the rates as parameterized in
\texttt{NUC123}~\cite{Kawano:1992ua},
\texttt{PArthENoPE}~\cite{Pisanti:2007hk,Consiglio:2017pot,Gariazzo:2021iiu}
and \texttt{AlterBBN}~\cite{Arbey:2011nf,Arbey:2018zfh} as well as to
the rate corresponding to the parameterization of the $S$-factor of
Eq.~(\ref{eq:p7Be8Bg_par}) in Fig.~\ref{fig:p7Be8Bg_rate}.

\subsection{The $\nuc{3}{H} + \nuc{4}{He} \to \nuc{7}{Li} + \gamma$ reaction}

In this case $Z_v = 1, Z_c=2$\,, $s_v = 1/2\,, s_c=0$ and $Z_f=3$ with
$J_f = {3}/{2}$ for the ground state and $J_f = {1}/{2}$ for the first
excited state.  The radiative capture cross section is determined by
electric dipole contributions only, \textit{i.e.}  by the expression
for the electric dipole contribution in
Eqs.~(\ref{eq:csE1},\ref{eq:amplitudesp}) with $k_C \ne 0$.  The
parameter sets labeled ``fit'' and ``A'' correspond to the parameter
sets labeled ``$\chi^2$'' and ``Model A'' in
Ref.~\cite{Premarathna:2019tup}, respectively.

\begin{figure}[!htb]
  \centering
  \includegraphics[width=0.475\textwidth,trim= 140 430 100 130, clip]{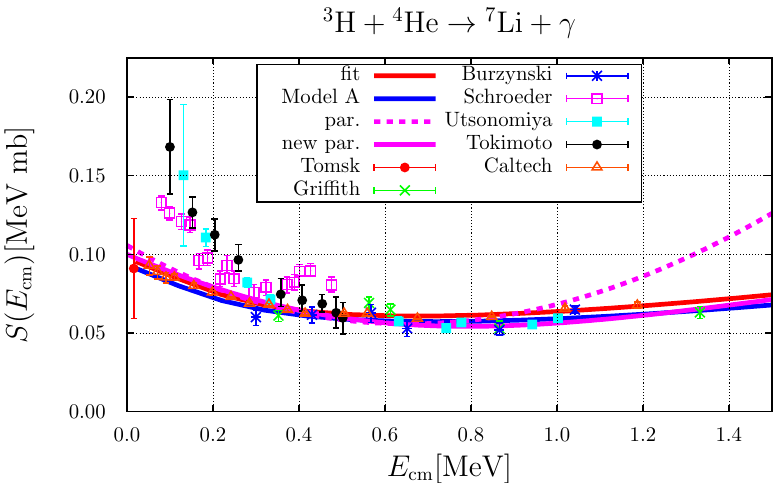} 
  \caption{$S$-factor as a function of the centre-of-mass energy 
    compared to experimental data:
    Tomsk~\cite{Bystritsky:2017gtz},
    Griffith~\cite{Griffith:1987xyz},
    Burzynski~\cite{Burzynski:1987bic},
    Schroeder~\cite{Schroder:1987opz},
    Utsonomiya~\cite{Utsunomiya:1988yqx,Utsunomiya:1990trd,Utsunomiya:1990zz,Utsunomiya:1992xyz},
    Tokimoto~\cite{Tokimoto:2001rd},
    Caltech~\cite{Brune:1994zz}.
    The blue and red curve correspond to the parameter sets ``A'' and
    ``fit'', respectively.  The dotted purple curve represents the
    $S$-factor from the parameterization as used in
    Ref.~\cite{Meissner:2023voo}\,.  The solid purple curve corresponds to
    the improved parameterization given in
    Eq.~(\ref{eq:3H4He7Lig_newpar})\,.
  }
  \label{fig:3H4He7Lig_S}
\end{figure}

The $S$-factor for this reaction
\begin{eqnarray}
  \label{eq:3H4He7Lig_newpar}
     S(E)
     &=&
     0.01\,\frac{1-1.15\,E+1.0\,E^2}{1+0.01\,E+0.5\,E^2}
        \nonumber\\
  &&
  \textnormal{(in MeV mb, with $E$ in MeV)}\,.
\end{eqnarray}
is compared to experimental data
and to the parameterization used in Ref.~\cite{Meissner:2023voo}
as well as  an improved parameterization, see
Eq.~(\ref{eq:3H4He7Lig_newpar}), thereof
in Fig.~\ref{fig:3H4He7Lig_S}.

\begin{figure}[!htb]
  \centering
  \includegraphics[width=0.475\textwidth,trim= 140 430 100 130, clip]{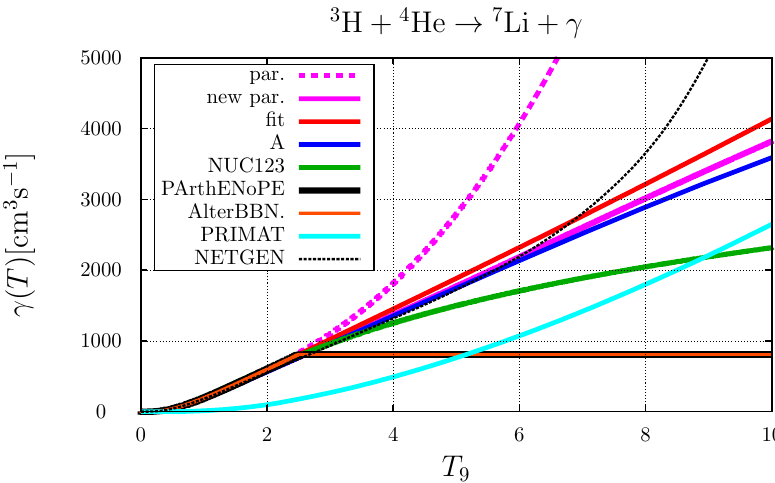} 
  \caption{
    Comparison of temperature
    ($T_9 = T / 10^9~\textnormal{K}$) dependent rates for the
    $\nuc{3}{H}+\nuc{4}{He} \to \nuc{7}{Li} + \gamma$ reaction.
    The blue and red curve represents the parameter sets ``A'' and
    ``fit'', respectively, the purple curve is the result corresponding to
    the parameterization of Eq.~(\ref{eq:3H4He7Lig_newpar}), while the
    dotted purple curve corresponds to the parameterization of
    Ref.~\cite{Meissner:2023voo}. Also shown are the parameterization of
    the rate as originally implemented in
    \texttt{NUC123}~\cite{Kawano:1992ua} (green curve),
    \texttt{PArthENoPE}~\cite{Pisanti:2007hk,Consiglio:2017pot,Gariazzo:2021iiu} (black curve),
    \texttt{AlterBBN}~\cite{Arbey:2011nf,Arbey:2018zfh} (brown curve) and
    \texttt{PRIMAT}~\cite{Pitrou:2018cgg} (cyan curve) as well as the parameterization provided by
    \texttt{NETGEN}, see~\cite{Xu:2012uw}.
    }
  \label{fig:3H4He7Lig_R} 
\end{figure}

This new parameterization of the $S$-factor is closer to the
calculations within the framework of Halo-EFT studied here and improves
the desciption of the data,
in particular for energies $E_{\textnormal{cm}}>1~\textnormal{MeV}$,
and indeed
yields a rate that is much smaller at higher temperatures, see Fig.~\ref{fig:3H4He7Lig_R}. 

\subsection{The $\nuc{3}{He} + \nuc{4}{He} \to \nuc{7}{Be} + \gamma$ reaction}

In this case $Z_v = 2, Z_c=2$\,, $s_v = 1/2\,, s_c=0$ and $Z_f=4$ with
$J_f = {3}/{2}$ for the ground state and $J_f = {1}/{2}$ for the first
excited state.  The radiative capture cross section is again
determined by electric dipole contributions only, \textit{i.e.}  by
the expression for the electric dipole contribution in
Eqs.~(\ref{eq:csE1},\ref{eq:amplitudesp}) with $k_C \ne 0$.
  The parameter sets labeled ``fit'' and ``AII'' correspond to
  the parameter sets labeled ``$\chi^2$'' and ``Model AII''
  in Ref.~\cite{Premarathna:2019tup}, respectively. 

\begin{figure}[!htb]
  \centering
  \includegraphics[width=0.475\textwidth,trim= 140 430 100 130, clip]{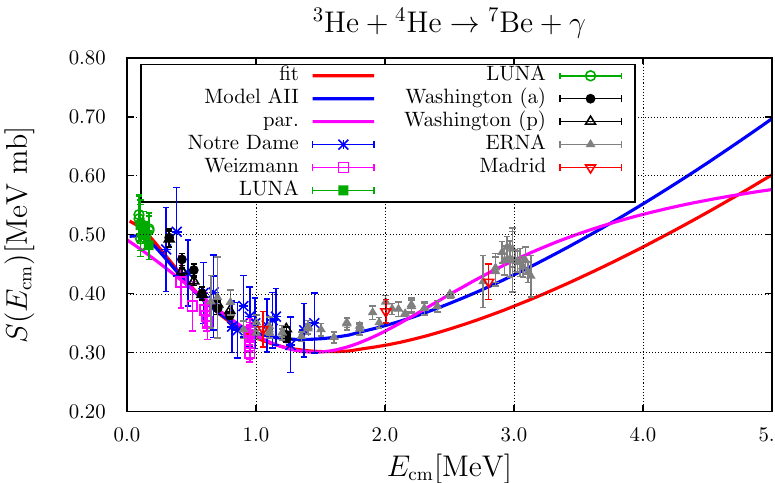} 
  \caption{$S$-factor as a function of the centre-of-mass energy
    compared to experimental data:
    Notre Dame~\cite{Kontos:2013qoa},
    Weizmann~\cite{NaraSingh:2004vj},
    LUNA~\cite{Gyurky:2007qq,LUNA:2007ffz},
    Washington~\cite{Brown:2007sj},
    ERNA~\cite{DiLeva:2009zz},
    Madrid~\cite{Carmona-Gallardo:2012apk}.
    The blue and red curve correspond to the parameter sets ``A II'' and ``fit'', respectively.
    The purple curve represents the parameterization as used in Ref.~\cite{Meissner:2023voo}\,.
    }
  \label{fig:3He4He7Beg_S}
\end{figure}

\begin{figure}[!htb]
  \centering
  \includegraphics[width=0.475\textwidth,trim= 140 430 100 130, clip]{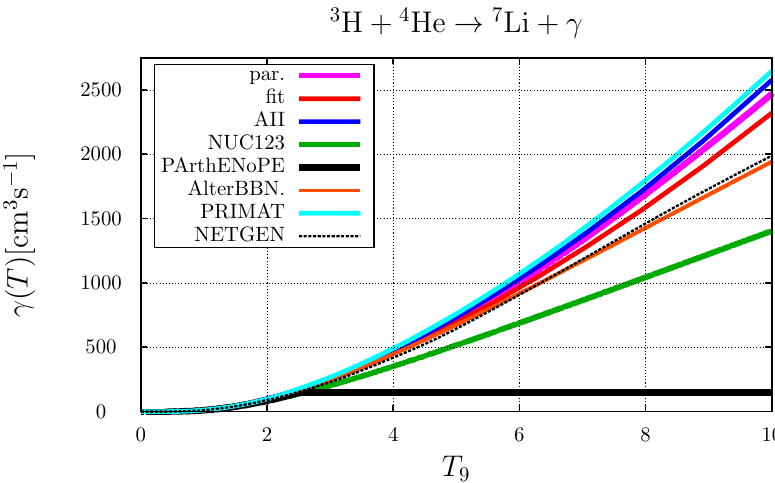}  
  \caption{Comparison of temperature ($T_9 = T / 10^9~\textnormal{K}$)
    dependent rates for the
    $\nuc{3}{He}+\nuc{4}{He} \to \nuc{7}{Be} + \gamma$ reaction.
    The blue and red curve represent the parameter sets ``AII'' and
    ``fit'', respectively, the purple curve corresponds to the parameterization of
    Ref.~\cite{Meissner:2023voo}. Also shown are the parameterizations of
    the rate as originally implemented in
    \texttt{NUC123}~\cite{Kawano:1992ua} (green curve),
    \texttt{PArthENoPE}~\cite{Pisanti:2007hk,Consiglio:2017pot,Gariazzo:2021iiu} (black curve),
    \texttt{AlterBBN}~\cite{Arbey:2011nf,Arbey:2018zfh} (brown curve) and
    \texttt{PRIMAT}~\cite{Pitrou:2018cgg} (cyan curve) as well as the parameterization provided by
    \texttt{NETGEN}, see~\cite{Xu:2012uw}.
    }
  \label{fig:3He4He7Beg_R} 
\end{figure}

The astrophysical $S$-factor is displayed in Fig.~\ref{fig:3He4He7Beg_S}.
The resulting nominal rates are given in Fig.~\ref{fig:3He4He7Beg_R}.

\section{\label{sec:alphadep}The fine-structure constant dependence of the rates}

In order to study the fine-structure constant dependence we calculated the rates for
\begin{equation}
  \label{eq:alhadep}
  \alpha = \alpha_0\,(1+\delta)\,,
\end{equation}
and the fractional change in $\alpha$, \textit{i.e.} $\delta$ was
varied in the range $[-0.05,+0.05]$\,.  We shall distinguish direct
and indirect effects of the variation of the fine-structure constant:

\paragraph{\label{sec:de}Direct effect}
First of all the fine-structure constant $\alpha$ enters the
calculation of the radiative capture cross section as a linear factor
due to the coupling of electromagnetic field to the charges and
currents, which in the amplitude is proportional to $e$ and hence in
the cross section leads to a proportionality $e^2 \propto \alpha$.
Furthermore $\alpha$ enters the cross section via the inverse 
Bohr-radius $k_C = Z_v\,Z_c c^2\,\mu\,\alpha$\,, that in turn determines
the dimensionless quantities $\eta_\gamma = k_C/\gamma$, where
$\gamma$ is the binding momentum, $\eta_\rho = k_C/\rho$, where
$\rho$ the $p$-wave effective range, and $\eta_p = k_C/p$ (the 
Sommerfeld-parameter), that enter the expressions for the
normalization $\mathcal{N}(\eta_\gamma,\eta_\rho)$
(Eq.~(\ref{eq:norm})), the amplitudes
$\mathcal{A}(\eta_\gamma;\eta_p)$ (Eq.~(\ref{eq:amplitudeAs}))\,, via
$X(\eta_\gamma;\eta_p)$ (Eq.~(\ref{eq:amplitudeX})), and
$\mathcal{B}(\eta_\gamma;\eta_p)$ (Eq.~(\ref{eq:BIntgndp}))\,, as well
as $Y(\eta_\gamma;\eta_p)$ (Eq.~(\ref{eq:amplitudeY}))\,.  The 
Sommerfeld-parameter $\eta_p$ also enters the astrophysical
$S$-factor.

Because the dependence of $k_C$ on $\alpha$ is linear, $k_C \propto
\alpha$, we have 
\begin{equation}
  \label{eq:kcalpha}
  k_C(\alpha)
  =
  k_C(\alpha_0\,(1+\delta_\alpha))
  =
  k_C(\alpha_0)\,(1+\delta_\alpha)\,.
\end{equation}
We shall call this the ``direct effect''.

\paragraph{\label{sec:ie}Indirect effect}
On top of this, the value of $\alpha$ influences the nuclear binding
energies, \textit{i.e.} the $\alpha$ dependence of the nuclear mass of the
nuclide $i$ is given by
\begin{equation}
  \label{eq:nucmass}
  m^i(\alpha)
  =
  m^i_N + V^i_C\,(1+\delta_\alpha)
  =
  m^i + V^i_C\,\delta_\alpha\,,
\end{equation}
where $V^i_C$ denotes the (repulsive) Coulomb-energy contribution to the nuclear mass.
This in turn influences the $Q$-value of the reaction, \textit{i.e.}
\begin{equation}
  \label{eq:Qvalue}
  Q(\alpha)
  =
  m_v(\alpha) + m_c(\alpha) - M_f(\alpha) = B_f(\alpha)
\end{equation}
and thus the binding momentum
\begin{equation}
  \label{eq:bindingmomentum}
  \gamma(\alpha)
  =
  \sqrt{2\,\mu(\alpha) c^2\,B_f(\alpha)}\,.
\end{equation}

Concerning the kinematics of the reaction: For a given
CMS kinetic energy $E$, the CMS relative momentum in the entrance
channel $p$ and the CMS final photon momentum $k_\gamma$ are given by
\begin{eqnarray}
  \label{eq:sE}
  s(E)
  &=&
      (m_v + m_c + E)^2\,,
  \\
  \label{eq:pE}
  p(E)
  &=&
      \sqrt{\frac{\left(s(E)-(m_c-m_v)^2\right)\left(s(E)-(m_c+m_v)^2\right)}{4\,s(E)}}
  \nonumber\\
  &&
      \approx
      \sqrt{2\,\mu\,E}\,,
      \label{kph}
  \\
  k_\gamma(E)
  &=&
      \frac{s(E)-M_f^2}{\sqrt{4\,s(E)}}
      \approx Q + E
\end{eqnarray}
and thus all depend on $\alpha$.  Because in general (for
$\delta_\alpha \approx 0.1$) $\Delta V^i_C/m^i \approx
\mathcal{O}(10^{-4})$ the dependence of $M$ and $\mu$ on $\alpha$ is
expected to be rather small ($\Delta\mu/\mu \approx
\mathcal{O}(10^{-4})$), whereas the change in the $Q$-value can be
appreciable, $\Delta Q/Q \approx \mathcal{O}(10^{-1})$\,. Accordingly,
the effect of a variation of $\mu$ with a variation of $\alpha$ on the
value of $k_C=Z_v\,Z_c\,\mu c^2\,\alpha$ will be ignored.

We  call the total of these kinematical variations the ``indirect effect''. 

In Ref.~\cite{Meissner:2023voo}, we introduced an approximation to the
dependence of the rate on $\alpha$ by evaluating the effect on the
parameterized $S$-factor at an energy where the $S$-factor is supposed
to be maximal. For charged particle induced reactions this energy is
given by
\begin{equation}
  \label{eq:rateapprox}
  \overline{E}
  =
  \left(\frac{kT}{2}\right)^{\frac{2}{3}}\left(E_G^i\right)^{\frac{1}{3}}\,.
\end{equation}
This then leads to a temperature dependent factor, that gives a fair
approximation to parameterized results, both with and without
including the indirect effects.

\begin{figure}[!htb]
  \centering
  \includegraphics[width=0.475\textwidth,trim= 140 430 100 130, clip]{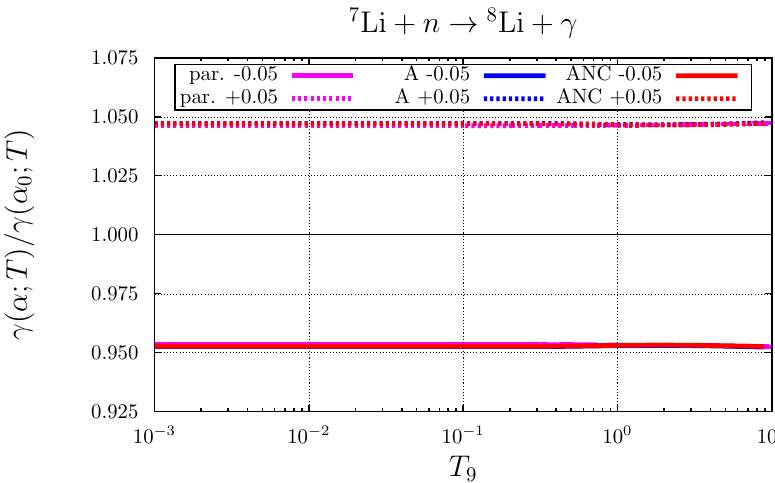}
  \includegraphics[width=0.475\textwidth,trim= 140 430 100 150, clip]{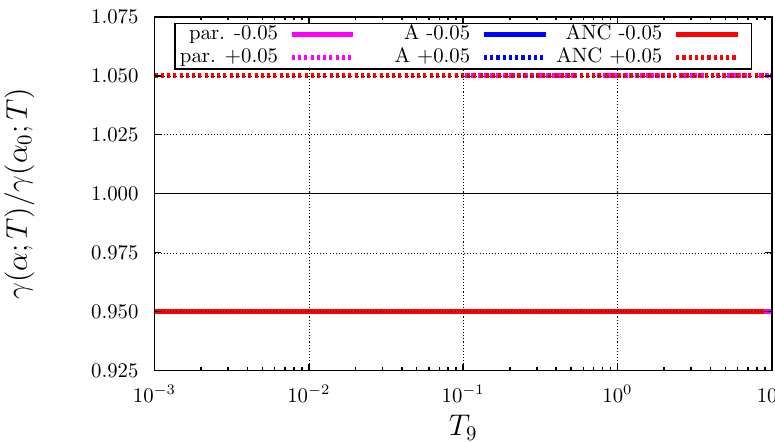}  
  \caption{
    Variation of the temperature ($T_9 = T / 10^9~\textnormal{K}$) dependent rate for the
    $n+\nuc{7}{Li} \to \nuc{8}{Li} + \gamma$ reaction with $\alpha=\alpha_0(1+\delta_\alpha)$.
    The solid curves correspond to $\delta=-0.05$, the dotted lines to $\delta=+0.05$\,.
    Top panel: including both the direct and indirect effects;
    bottom panel: including the direct effect only.
  }
  \label{fig:n7Li8Lig_rate_rel}
\end{figure}

\subsection{The $n + \nuc{7}{Li} \to \nuc{8}{Li} + \gamma$ reaction}

Since the neutron in the entrance channel is uncharged, there is no Coulomb
interaction between the clusters and accordingly the direct effect of
varying $\alpha$-dependence is completely determined by the fact that the cross section
is strictly linear in $\alpha$. In addition there is the indirect effect
stemming from the fine-structure constant dependence of the Coulomb
contributions to the bindings energies of $\nuc{7}{Li}$ and
$\nuc{8}{Li}$, that affects the $Q$-value of the reaction.

The variation of the rate with $\alpha$ is displayed in
Fig.~\ref{fig:n7Li8Lig_rate_rel}, where the relative variation
$\gamma(\alpha;T)/\gamma(\alpha_0;T)$ is plotted as a function of the
temperature. Although the calculated rates, see
Fig.~\ref{fig:n7Li8Lig_rate}, do differ slightly the relative changes
of the rates are almost identical. The bottom panel in Fig.~\ref{fig:n7Li8Lig_rate_rel} indeed
merely reflects that the cross section for this reaction trivially
linearly depends on $\alpha$\,, \textit{i.e.} if $\alpha$ varies by
5\%, then also the direct effect rate varies by 5\% and this effect is
temperature-independent. Small deviations occur if also the variation
of the binding energies of the Li-nuclides is taken into account, see
the top panel of Fig.~\ref{fig:n7Li8Lig_rate_rel}.

\subsection{The $p + \nuc{7}{Be} \to \nuc{8}{B} + \gamma$ reaction}

\begin{figure*}[!htb]
  \centering
  \includegraphics[width=0.475\textwidth,trim= 140 430 90 130, clip]{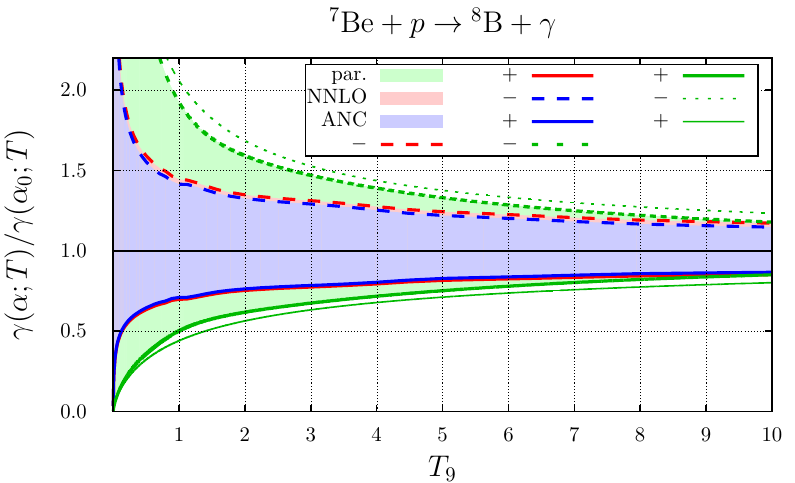}
  \includegraphics[width=0.475\textwidth,trim= 140 430 90 130, clip]{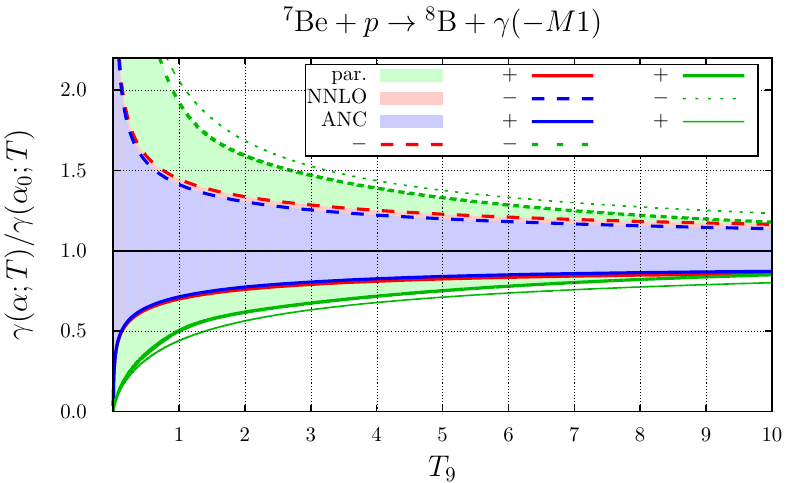}
  \\
  \includegraphics[width=0.475\textwidth,trim= 140 430 90 130, clip]{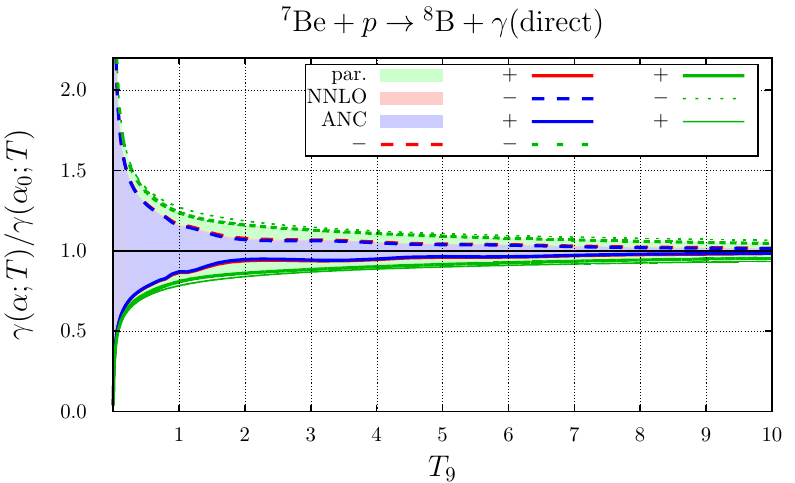}
  \includegraphics[width=0.475\textwidth,trim= 140 430 90 130, clip]{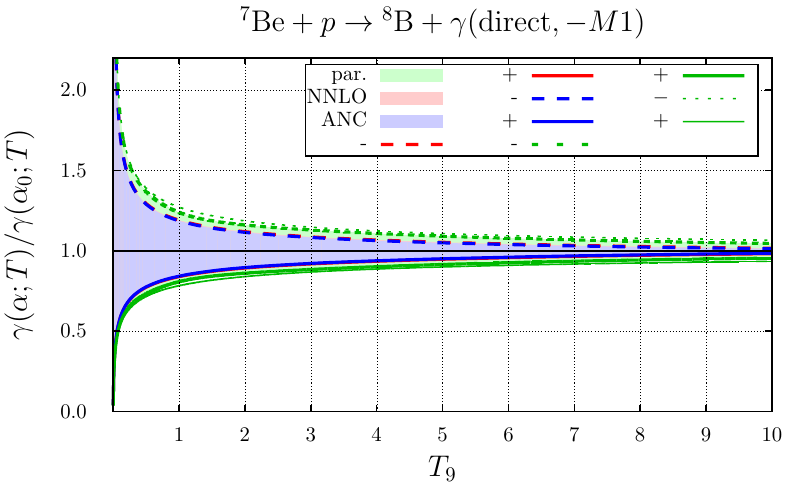}
  \caption{
    Variation of the rate normalized to the rate at the nominal
    $\alpha_0$ with varying $\alpha$, \textit{i.e.}
    $\gamma(T_9;\alpha_0(1+\delta_\alpha))/\gamma(T_9;\alpha_0)$ for the
    $p+\nuc{7}{Be} \to \nuc{8}{B} + \gamma$ reaction.
    Here, $T_9 = T / 10^9~\textnormal{K}$.
    Top left: Total rate, including the $M1$ contribution.
    Top right: Total rate, without the $M1$ contribution.
    Bottom left: Direct variation only, including the $M1$ contribution.
    Bottom right: Direct variation only, without the $M1$ contribution.
    The coloured areas bounded by the curves correspond to a variation $\delta \in [-0.05,+0.05]$\,.
    The green curves (areas) correspond to the variation of the rates
    with $\alpha$ according to the penetration factor and the trivial
    linear dependence of the cross section on $\alpha$ for dominant dipole
    radiation (direct effects). The blue and red curves (areas)
    represent the results for the parameter sets ``ANC'' and ``NNLO''\,.
    The solid lines (marked '+') correspond to $\alpha=0.05$, the dotted lines to $\alpha=-0.05$.
    The thinner lines represent the results from the approximation to
      the fine-structure constant dependence of the rate based on the
      temperature-dependent factor evaluated at the energy $\overline{E}$
      given in Eq.~(\ref{eq:rateapprox}).
  } 
  \label{fig:p7Be8Bg_rate_rel}
\end{figure*}

The  fine-structure dependence of the temperature dependent rate of the
proton-induced radiative capture reaction is more interesting. In
Fig.~\ref{fig:p7Be8Bg_rate_rel} the variation of the temperature-dependent
rate with $\alpha$ of the calculated values with the two parameter
sets is compared to the values obtained with the parameterization of
the $\alpha$-dependence of the rates based on the parameterized cross
sections as done in Ref.~\cite{Meissner:2023voo}. From this figure one
infers that the relative variation in the Halo-EFT calculations is
smaller by about $40\%$ than that found in Ref.~\cite{Meissner:2023voo}.
Excluding the $M1$ contribution yields practically identical results.
Furthermore it is observed, that considering also the indirect effect,
\textit{i.e.} also the effect on the binding energies and thus on the
$Q$-value of the reaction, enhances this difference.

\subsection{The $\nuc{3}{H} + \nuc{4}{He} \to \nuc{7}{Li} + \gamma$ reaction}

\begin{figure}[!htb]
  \centering
  \includegraphics[width=0.45\textwidth,trim=140 430 90 130, clip]{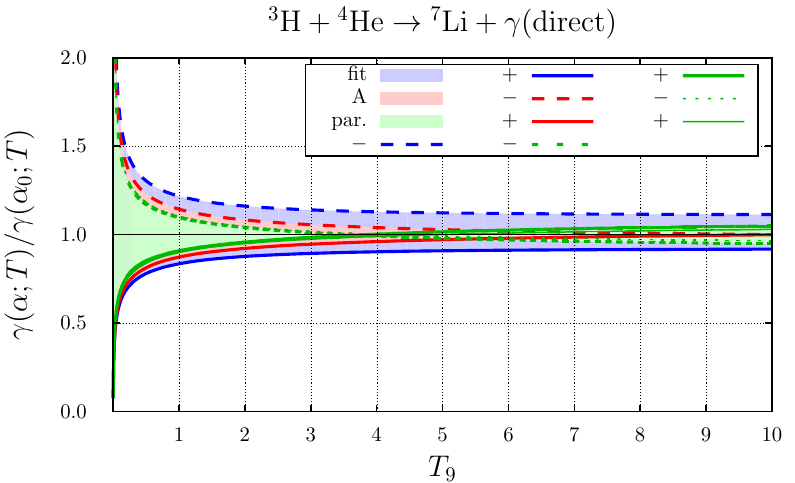} 
  \includegraphics[width=0.45\textwidth,trim=140 430 90 130, clip]{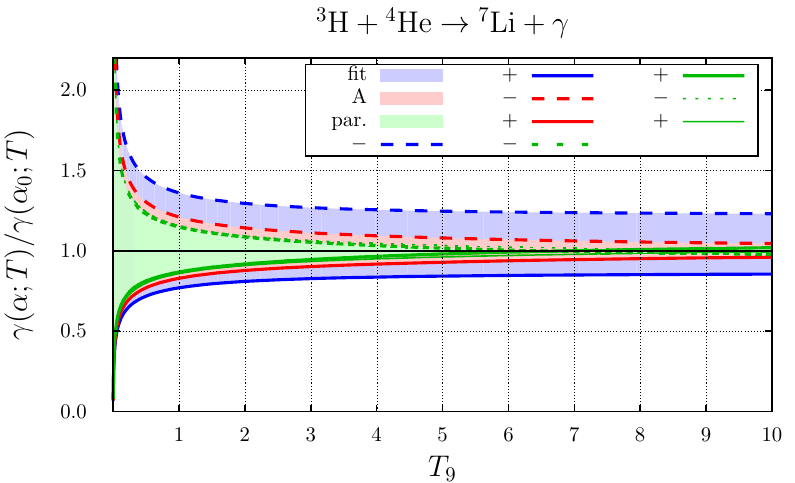}
  \caption{
    Fine-structure constant dependence of the temperature dependent
    rate of the $\nuc{3}{H}+\nuc{4}{He}\to\nuc{7}{Li}+\gamma$
    reaction. Shown is the rate relative to rate with the nominal value:
    ``+'' is the rate at $\alpha=1.05\,\alpha_0$\,, ``-'' the value at
    $\alpha=0.95\,\alpha_0$\,.
    Here, $T_9 = T / 10^9~\textnormal{K}$.
    The blue ranges were obtained with the
    parameter set ``fit'', the red range with the parameter set ``A''
    and the green range represents the variation of $\alpha$ as determined
    in Ref.~\cite{Meissner:2023voo}\,.
    The solid lines (marked '+') correspond to $\alpha=0.05$, the dotted lines to $\alpha=-0.05$.
     The fine-structure constant dependence of the rate based on the
      temperature dependent factor evaluated at the energy $\overline{E}$
      given in Eq.~(\ref{eq:rateapprox}).
  } 
  \label{fig:rate7Lirel}
\end{figure}

The relative variation with $\alpha$ of the temperature-dependent rate
for the two parameter sets ``fit'' and ``A'' are compared to that with
the rate based on the parameterization of Ref.~\cite{Meissner:2023voo}
in Fig.~\ref{fig:rate7Lirel}.  Contrary to the previous reaction this
variation is larger for the Halo-EFT results, in particular for the
parameter set ``fit'' than that with the parameterized rate.
Considering the direct effect only leads to the same conclusion.

\subsection{The $\nuc{3}{He} + \nuc{4}{He} \to \nuc{7}{Be} + \gamma$ reaction}

\begin{figure}[!htb] 
  \centering
  \includegraphics[width=0.45\textwidth,trim=140 430 90 130, clip]{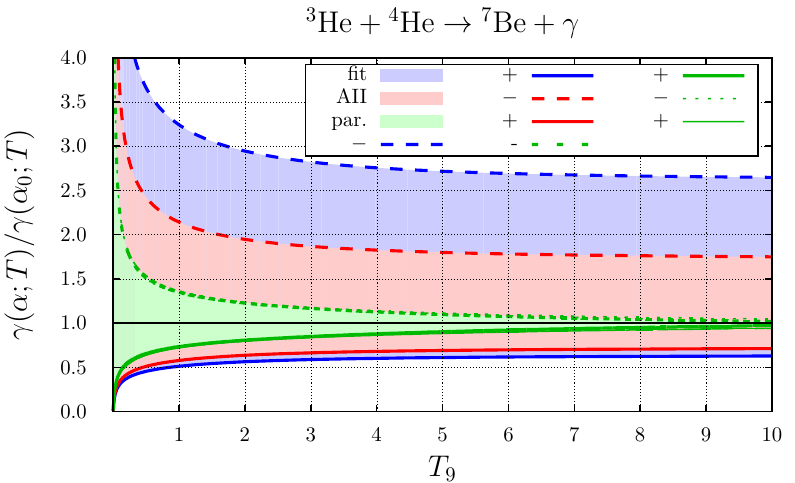}
  \includegraphics[width=0.45\textwidth,trim=140 430 90 130, clip]{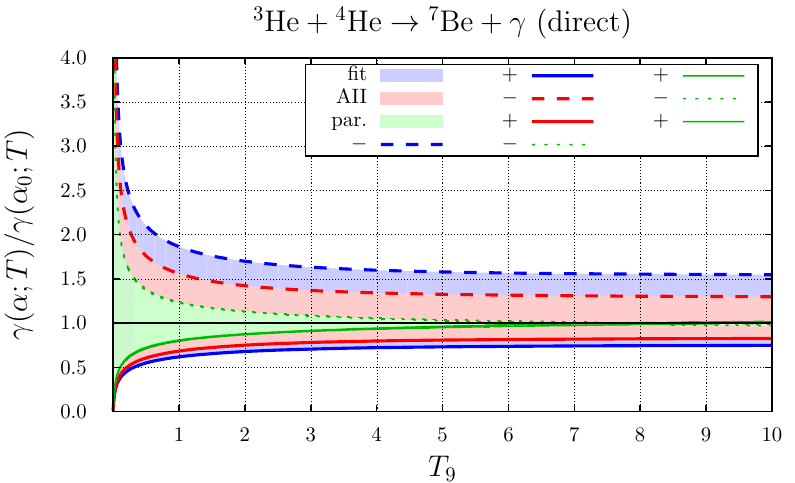}
  \caption{ 
    Fine-structure constant dependence of the temperature dependent
    rate of the $\nuc{3}{He}+\nuc{4}{He}\to\nuc{7}{Be}+\gamma$
    reaction. Here, $T_9 = T / 10^9~\textnormal{K}$.
    Shown is the rate relative to rate with the nominal value:
    ``+'' is the rate at $\alpha=1.05\,\alpha_0$\,, ``-'' the value at
    $\alpha=0.95\,\alpha_0$\,. The blue ranges were obtained with the
    parameter set ``fit'', the red range with he parameter set ``A''
    and the green range represents the variation of $\alpha$ as determined
    in Ref.~\cite{Meissner:2023voo}\,.
    The solid lines (marked '+') correspond to $\alpha=0.05$, the dotted lines to $\alpha=-0.05$.
    The fine-structure constant dependence of the rate based on the
      temperature-dependent factor evaluated at the energy $\overline{E}$
      given in Eq.~(\ref{eq:rateapprox}).
  }  
  \label{fig:rate7Berel} 
\end{figure}

The relative variation of the rate with the value of the
fine-structure constant $\alpha$ of the temperature-dependent rates
for the two parameter sets ``fit'' and ``AII'' are compared to that
with the rate based on the parameterization of
Ref.~\cite{Meissner:2023voo} in Fig.~\ref{fig:rate7Berel}.  Here, the
relative variation with the two Halo-EFT parameter sets is much larger
than for the parameterization used previously, in particular if the
fine-structure constant is smaller than the nominal value.  We shall
discuss the reason for this in the next section,
Sect.~\ref{sec:discussion}.

\begin{figure*}[!htb]
  \centering
  \includegraphics[width=0.3\textwidth,trim= 140 320 90 130, clip]{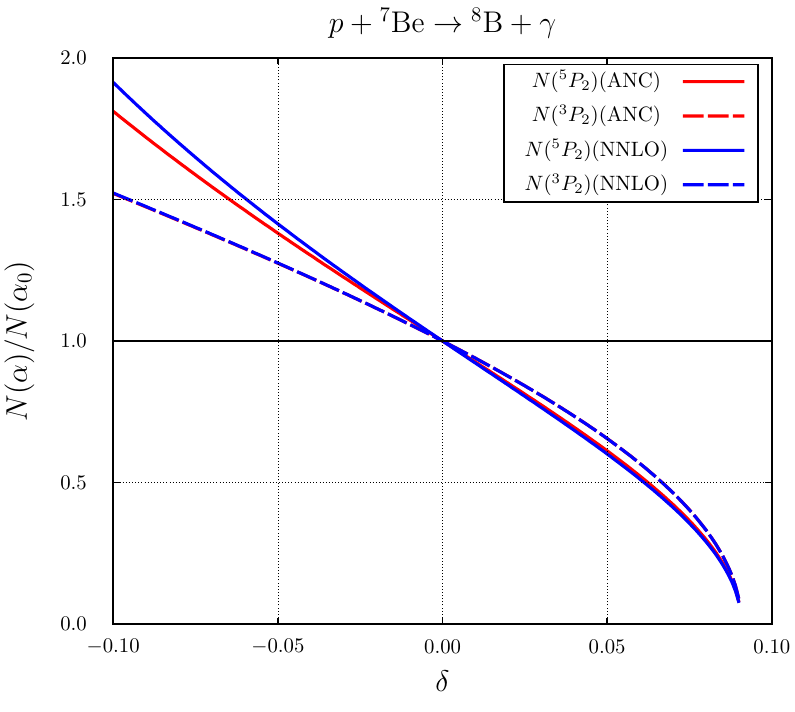}
  \includegraphics[width=0.3\textwidth,trim= 140 320 90 130, clip]{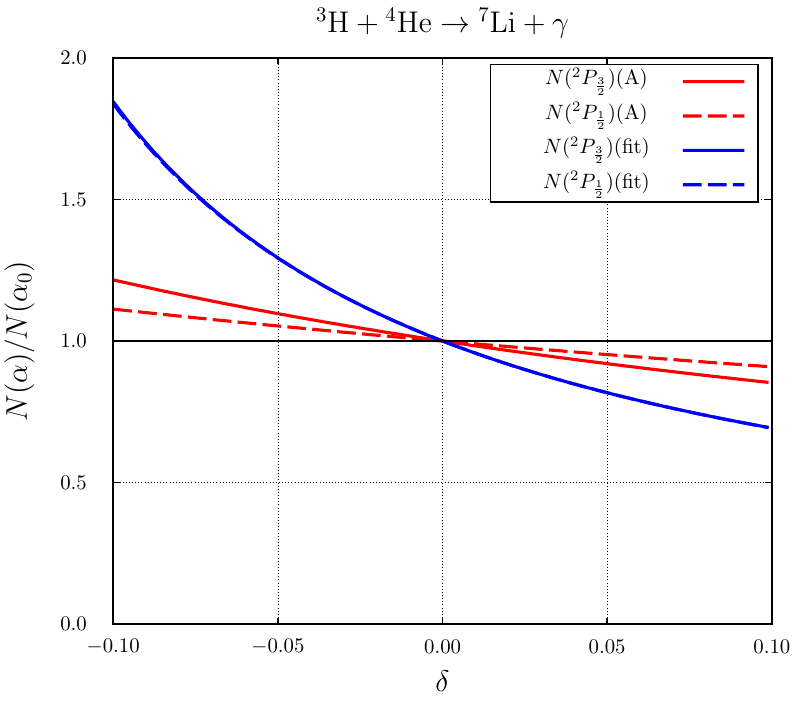}
  \includegraphics[width=0.3\textwidth,trim= 140 320 90 130, clip]{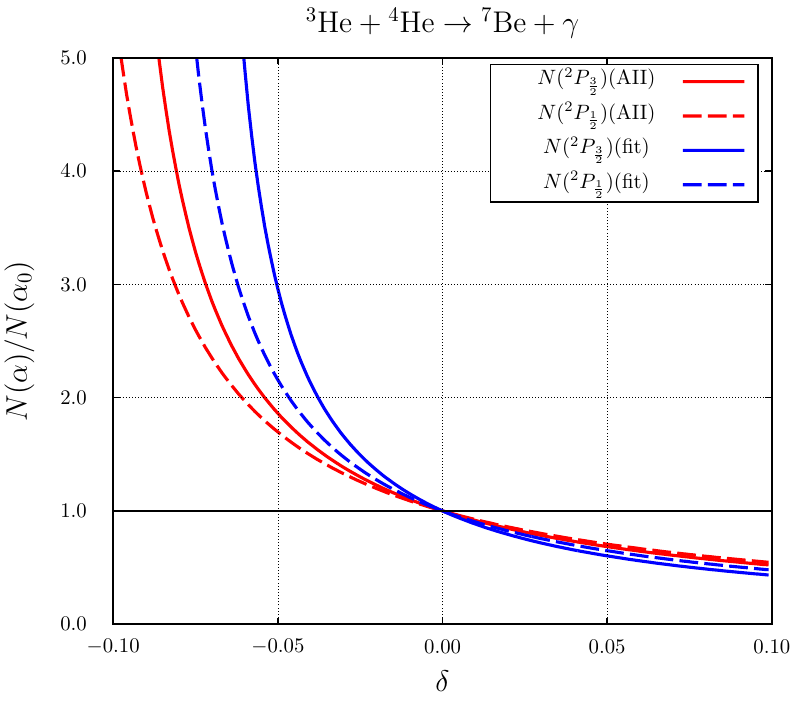}
  \caption{
    Variation of the relative normalization
    $N(\alpha)/N(\alpha_0)$ with $\alpha = \alpha_0(1+\delta)$ for the reactions
    $p+\nuc{7}{Be} \to \nuc{8}{B} + \gamma$ (left),
    $\nuc{3}{H}+\nuc{4}{He} \to \nuc{7}{Li} + \gamma$ (middle),
    $\nuc{3}{He}+\nuc{4}{He} \to \nuc{7}{Be} + \gamma$ (right).
    For the first reaction the results are shown for both the $\spc{5}{P}{2}$ and the $\spc{3}{P}{2}$
    amplitudes contributing with equal weight to the ground state capture. For
    the other two reactions the normalisation of the
    $\spc{2}{P}{\frac{3}{2}}$ (ground state) and the
    $\spc{2}{P}{\frac{1}{2}}$ (excited state) are shown.
    The first reaction ceases to be exo-energetic for positive $\delta \approx
    0.1$. For the last reaction the expression for the normalisation, see
    Eq.~(\ref{eq:norm}), exhibits a pole for negative $\delta \approx
    -0.06$.
  } 
  \label{fig:relnorm}
\end{figure*}

\subsection{\label{sec:discussion}Discussion}

First of all we observe that although the resonant magnetic dipole
contribution in the nucleon induced reaction accounts for a prominent
feature in the cross sections (or astrophysical $S$-factors), this
contribution is of minor importance in the variation of the rates with
$\alpha$, see \textit{e.g.} Fig.~\ref{fig:p7Be8Bg_rate_rel}.  Below we
shall therefore focus on the effects from the dominant electric dipole
contributions.  With the exception of the neutron induced reaction the
effects of the $\alpha$-variation differ from what was estimated on
the basis of the parameterization of the cross-section used before in
Ref.~\cite{Meissner:2023voo}.  The dominant effect from the electric
dipole contribution seems to be the $\alpha$ variation of the
normalisation of Eq.~(\ref{eq:norm}).  The variation with $\alpha$ of
the relative normalisation $N(\alpha)/N(\alpha_0)$ for the three
charged particle induced reactions is displayed in
Fig.~\ref{fig:relnorm}. For the proton induced reaction the results
are shown for both the $\spc{5}{P}{2}$ and the $\spc{3}{P}{2}$
amplitudes contributing with equal weight to the ground state
capture. For the other two reactions the normalisation of the
$\spc{2}{P}{\frac{3}{2}}$ (ground state) and the
$\spc{2}{P}{\frac{1}{2}}$ (excited state) are shown.  This figure
illustrates the main effects observed in the variation of the rates
with $\alpha$:
\begin{enumerate}
\item
  Because of the absence of Coulomb interactions the variation with
  $\alpha$ of the cross sections and corresponding rates of the neutron
  induced reaction is trivially linear.
\item
  For the proton induced reaction the normalisation varies with
  $\alpha$ almost linearly by $\pm 40\%$ for $\delta \in [-0.05,0.05]$.
  Furthermore, the results for the two parameter sets considered here
  are almost identical. The variation of the rates is slightly larger
  for negative $\delta$, while considering the direct effect alone leads to a
  variation symmetric in $\delta$, in accordance with the variation of the
  normalisation. 
  Also displayed in Fig.~\ref{fig:p7Be8Bg_rate_rel} is
  the variation of the parameterized rate with $\alpha$ on the basis of
  the approximation introduced in Ref.~\cite{Meissner:2023voo}, by evaluating
  the effects at a fixed energy, see also Eq.~(\ref{eq:rateapprox}).
  Indeed for the proton induced
  reaction this temperature dependence is larger than the result
  calculated in Halo-EFT in both cases.
\item
  In case of the $\nuc{3}{H}$ and $\nuc{3}{He}$ induced reactions the
  results with the approximation discussed above almost coincide with the
  results on the basis of the parameterization, as was already
  demonstrated in Ref.~\cite{Meissner:2023voo} and indeed in these cases
  is much smaller than what is to expected on the basis of the
  normalisation. Anyhow, the $\alpha$ dependence of the normalisation is
  rather asymmetric in $\delta$ as shown in
  Fig.~\ref{fig:relnorm}. For the $\nuc{3}{He}$ induced reaction
  this is even more prominent since the
  denominator in the expression for the norm, see Eq.~(\ref{eq:norm}),
  vanishes for $\alpha < -0.06$,
  corresponding to a pole in the normalisation and thus leading to a
  very asymmetric $\delta$ dependence in this case.
\item
  Accordingly, within Halo-EFT the study of the $\alpha$ dependence of the
  rates is limited to a rather
  moderate relative variation of $\alpha$ of $5\%$ only. 
\end{enumerate}

\section{\label{sec:abun}Abundances}

  In order to assess the relevance of the variation of the rates with
a variation of the fine-structure constant on the variation of the
resulting abundances of the light elements in BBN, we used five
different publicly available codes, \textit{viz.}
\texttt{NUC123}~\cite{Kawano:1992ua}, 
\texttt{PArthENoPE}~\cite{Gariazzo:2021iiu}, 
\texttt{AlterBBN}~\cite{Arbey:2018zfh},  
\texttt{PRIMAT}~\cite{Pitrou:2018cgg} and
\texttt{PRyMordial}~\cite{Burns:2023sgx,PRyMordial-Code}.
We use the rates as in our previous work,
see~\cite{Meissner:2023voo}, substituting the $\alpha$ dependence of
the rates for the four reactions considered here as discussed above.
More specifically, for calculating the $\alpha$ dependence of the
abundances, we used the parameter sets ``NNLO'', ``fit'' and ``fit''
for the
$p + \nuc{7}{Be} \to \nuc{8}{B} + \gamma$,
$\nuc{3}{H} + \nuc{4}{He} \to \nuc{7}{Li} + \gamma$ and 
$\nuc{3}{He} + \nuc{4}{He} \to \nuc{7}{Be} + \gamma$ reactions,
respectively, the $\alpha$-dependence of neutron induced radiative
capture reaction $n + \nuc{7}{Li} \to \nuc{8}{Li} + \gamma$ being
practically linear anyway. As was demonstrated in the
Sect.~\ref{sec:alphadep} these parameter sets showed the largest
variation of the rates with $\alpha$.

\begin{table}[hbt]
  \begin{ruledtabular}
    \caption{\label{tab:nomabun}%
      Nominal abundances as number ratios $Y_n/Y_H$ (for
      ${}^4\textnormal{He}$ the mass ratio $Y_p$) calculated with the
      modified versions of the codes as in Ref.~\cite{Meissner:2023voo}, but
      with the nominal results (\textit{i.e.} $\alpha=\alpha_0$) for the
      four reactions considered in this work.  The value of the
      baryon-to-photon ratio and the nominal value of the neutron lifetime
      are $\eta = 6.14 \cdot 10^{-10}$ and $\tau_n = 879.4\,\textnormal{s}$,
      respectively.  For comparison also the values previously obtained in
      Ref.~\cite{Meissner:2023voo} are listed.}
    \begin{tabular}{@{}l%
      @{\extracolsep{\fill}}d%
      @{\extracolsep{\fill}}d%
      @{\extracolsep{\fill}}d%
      @{\extracolsep{\fill}}d%
      @{\extracolsep{\fill}}d%
      @{}}
      \texttt{code}
      & \mc{$\nuc{2}{H}$}
      & \mc{$\nuc{3}{H}\!\!+\!\!\nuc{}{He}$}
      & \mc{$Y_{p}$}
      & \mc{$\nuc{6}{Li}$}
      & \mc{$\nuc{7}{Li}\!\!+\!\!\nuc{}{Be}$}
      \\[-0.25ex]
      & \mc{$\times 10^5$}
      & \mc{$\times 10^5$}
      & \mc{}
      & \mc{$\times 10^{14}$}
      & \mc{$\times 10^{10}$}
      \\
      \colrule
      \texttt{NUC123}        & 2.500 & 1.139 & 0.246 & 1.808 & 5.540 \\
      \cite{Meissner:2023voo}& 2.501 & 1.139 & 0.246 & 1.809 & 5.172 \\
      \texttt{PArthENoPE}    & 2.569 & 1.147 & 0.247 & 1.819 & 5.376 \\
      \cite{Meissner:2023voo}& 2.569 & 1.147 & 0.247 & 1.820 & 5.017 \\
      \texttt{AlterBBN}      & 2.585 & 1.153 & 0.248 & 1.903 & 5.350 \\
      \cite{Meissner:2023voo}& 2.585 & 1.153 & 0.248 & 1.904 & 4.993 \\
      \texttt{PRIMAT}        & 2.562 & 1.150 & 0.247 & 1.861 & 5.394 \\
      \cite{Meissner:2023voo}& 2.563 & 1.149 & 0.247 & 1.862 & 5.033 \\
      \texttt{PRyMordial}    & 2.581 & 1.148 & 0.247 & 1.891 & 5.448 \\
      \colrule
      PDG~\cite{Workman:2022ynf}  & 2.547 &       & 0.245 &       & 1.6 \\
      $\qquad\pm$  & 0.025 &       & 0.003 &       & 0.3 \\  
    \end{tabular}
  \end{ruledtabular}
\end{table}   

In Table~\ref{tab:nomabun} we list the nominal rates, \textit{i.e.}
for $\alpha=\alpha_0$, see Eq.~(\ref{eq:alpha0}). The results show
that with the exception of the values for the
$\nuc{7}{Li}\!\!+\!\!\nuc{}{Be}$ abundance, which are larger by about
$10\%$, and thus slightly deteriorate the so-called ``Li-problem'',
the treatment of the four reactions in Halo-EFT as considered here
leads to results practically identical to those obtained previously in
Ref.~\cite{Meissner:2023voo}.

\begin{figure*}[!htb]
  \centering
  \includegraphics[width=0.97\textwidth, trim=105 325 65 50, clip]{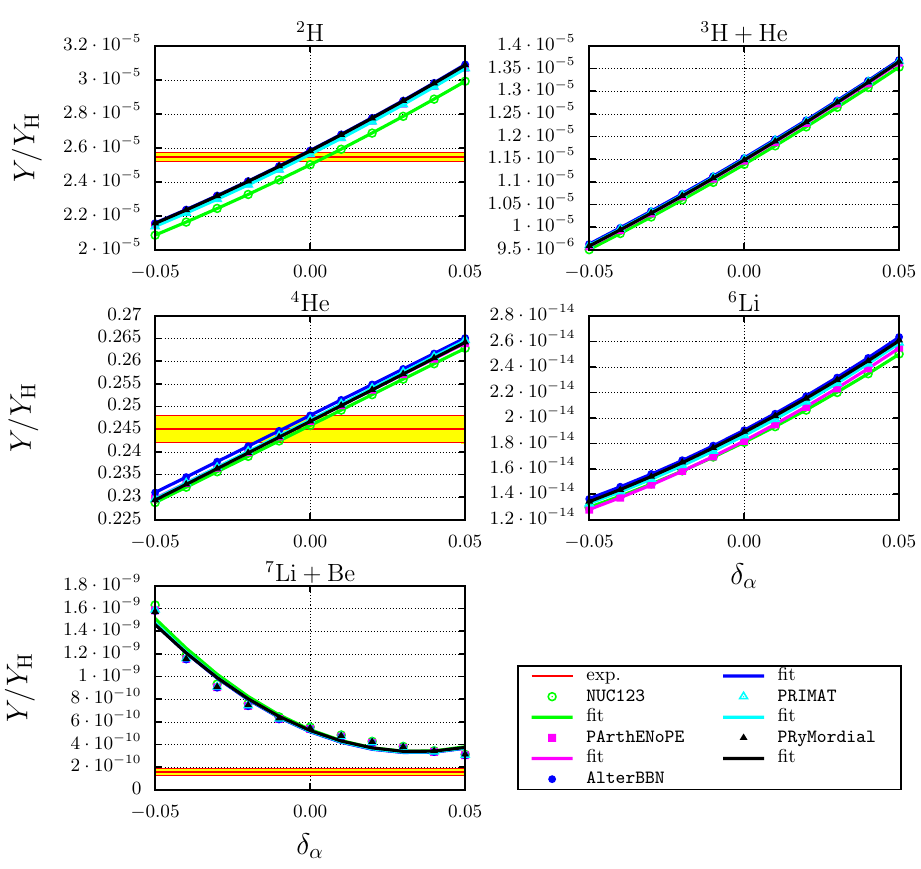}
  \caption{
    \label{fig:fabund}
    Variation of the abundance ratios $Y_n/Y_H$ with a variation of
    $\alpha = \alpha_0\,(1+\delta_\alpha)$ for $\delta_\alpha \in
    [-0.05,0.05]$ obtained with the codes:
    \texttt{NUC123}~\cite{Kawano:1992ua},
    \texttt{AlterBBN}~\cite{Arbey:2018zfh},
    \texttt{PArthENoPE}~\cite{Gariazzo:2021iiu},
    \texttt{PRIMAT}~\cite{Pitrou:2018cgg} and
    \texttt{PRyMordial}~\cite{Burns:2023sgx,PRyMordial-Code}.
    Here, we use
    $\eta=6.14 \cdot 10^{-10}$
    and
    $\tau_n = 879.4\,\textnormal{s}$\,.
    Also shown are the solid curves obtained by the fits according
    to Eq.~(\ref{eq:quadfit}) with the parameters listed in
    Table~\ref{tab:rme}.  The experimental values cited in
    PDG~\cite{Workman:2022ynf} (thick red lines) are indicated by
    yellow-highlighted regions (color online) representing the $1\sigma$
    limits by red lines.
  }
\end{figure*}

The fine-structure constant dependence
of the primordial abundances is depicted in Fig.~\ref{fig:fabund}.
 
Again, the results for the $d-$, $\nuc{3}{H}\!\!+\!\!\nuc{}{He}-$,
$\nuc{4}{He}-$ and $\nuc{6}{Li}-$ abundances are very similar to those
obtained previously, see Fig.~5 in
Ref.~\cite{Meissner:2023voo}. Moreover, the five BBN-codes considered
here produce consistent results, in spite of the fact that these codes
differ in details, such as the number of reactions in the BBN network
or the manner in which the rate equations are solved numerically.
This then also applies to the values for the resulting response matrix
elements. The (linear) response matrix elements
$\partial\log{(Y_n/Y_H)}/\partial\log{\alpha} = c_1$ and the
coefficients of the quad{\-}ratic term ($c_2$) in a quadratic
least-squares fit of the form
\begin{equation}
  \label{eq:quadfit}
  P_k(\delta_\alpha) = c_0\left(1+c_1\,\delta_\alpha^{} + c_2\,\delta_\alpha^2\right)\,,    
\end{equation}
are given and compared to the results obtained previously in
Table~\ref{tab:rme}. 

\begin{table*}[hbt]
  \caption{\label{tab:rme}%
    BBN response matrix $c_1
    = \partial\log{(Y_n/Y_H)}/\partial\log{\alpha}$ and the coefficients
    $c_2$ of the quadratic term in Eq.~(\ref{eq:quadfit}) at $\eta=6.14
    \cdot 10^{-10}$ and $\tau_n = 879.4\,\textnormal{s}$\,. $Y_n/Y_H$ are
    the number ratios of the abundances relative to hydrogen; $Y_p$ is
    conventionally the ${}^4$He/H mass ratio. The results obtained with
    the five BBN codes
    \texttt{NUC123}~\cite{Kawano:1992ua},
    \texttt{PArthENoPE}~\cite{Gariazzo:2021iiu},
    \texttt{AlterBBN}~\cite{Arbey:2018zfh},
    \texttt{PRIMAT}~\cite{Pitrou:2018cgg} and
    \texttt{PRyMordial}~\cite{Burns:2023sgx,PRyMordial-Code}
    are compared to the results previously obtained in
    Ref.~\cite{Meissner:2023voo}. The odd rows show the results with the
    rates for the reactions evaluated in the present contribution for the
    reactions
    $p + \nuc{7}{Be} \to \nuc{8}{B} + \gamma$,
    $\nuc{3}{H} + \nuc{4}{He} \to \nuc{7}{Li} + \gamma$ and 
    $\nuc{3}{He} + \nuc{4}{He} \to \nuc{7}{Be} + \gamma$ where the parameter sets
    ``NNLO'', ``fit'' and ``fit'' were used, respectively.
    Otherwise the rates are identical to those in ~\cite{Meissner:2023voo}.
  }
  \begin{ruledtabular}
    \begin{tabular}{@{}l%
        @{\extracolsep{2.0em}}d%
        @{\extracolsep{0.4em}}d%
        @{\extracolsep{2.0em}}d%
        @{\extracolsep{0.4em}}d%
        @{\extracolsep{2.0em}}d%
        @{\extracolsep{0.4em}}d%
        @{\extracolsep{2.0em}}d%
        @{\extracolsep{0.4em}}d%
        @{\extracolsep{2.0em}}d%
        @{\extracolsep{0.4em}}d%
        @{}} 
      \texttt{code}
      & \multicolumn{2}{c}{$\nuc{2}{H}$}
      & \multicolumn{2}{c}{$\nuc{3}{H}\!\!+\!\!\nuc{3}{He}$}
      & \multicolumn{2}{c}{$Y_{p}$}
      & \multicolumn{2}{c}{$\nuc{6}{Li}$}
      & \multicolumn{2}{c}{$\nuc{7}{Li}\!\!+\!\!\nuc{7}{Be}$}
      \\
      \cline{2-3}\cline{4-5}\cline{6-7}\cline{8-9}\cline{10-11}
      & \mc{$c_1$} & \mc{$c_2$}
      & \mc{$c_1$} & \mc{$c_2$}
      & \mc{$c_1$} & \mc{$c_2$}
      & \mc{$c_1$} & \mc{$c_2$}
      & \mc{$c_1$} & \mc{$c_2$}
      \\
      \colrule
      \texttt{NUC123}         &  3.620 & 6.198 & 3.539 & 4.661 & 1.386 & 0.035 & 6.642 & 20.074 & -21.296 & 312.074 \\
      \cite{Meissner:2023voo} &  3.655 & 6.228 & 3.540 & 4.625 & 1.387 & 0.016 & 6.830 & 20.412 & -4.325 & 7.480 \\
      \texttt{PArthENoPE}     &  3.606 & 6.173 & 3.534 & 4.619 & 1.390 & 0.056 & 6.968 & 21.116 & -21.284 & 312.328 \\
      \cite{Meissner:2023voo} &  3.635 & 6.182 & 3.533 & 4.577 & 1.389 & 0.065 & 7.159 & 21.482 & -4.308 & 7.715 \\
      \texttt{AlterBBN}       &  3.610 & 6.135 & 3.526 & 4.591 & 1.375 & 0.048 & 6.651 & 20.167 & -21.312 & 312.939 \\
      \cite{Meissner:2023voo} &  3.644 & 6.188 & 3.526 & 4.568 & 1.373 & 0.049 & 6.857 & 20.499 & -4.322 & 7.865 \\
      \texttt{PRIMAT}         &  3.627 & 6.253 & 3.535 & 4.631 & 1.415 & 0.072 & 6.754 & 20.593 & -21.273 & 311.595 \\
      \cite{Meissner:2023voo} &  3.658 & 6.264 & 3.534 & 4.595 & 1.408 & 0.081 & 6.953 & 20.828 & -4.302 & 7.563 \\
      \texttt{PRyMordial}     &  3.609 & 5.975 & 3.544 & 4.756 & 1.411 & 0.081 & 6.698 & 18.453 & -20.694 & 300.083 \\
    \end{tabular}
  \end{ruledtabular}
\end{table*}  

We do find a very different result for the $\alpha$-dependence of the
$\nuc{7}{Li}\!\!+\!\!\nuc{}{Be}$ abundance: In particular the linear
response coefficient is approximately five times larger than the value
obtained previously and moreover the response is far from linear, the
quadratic coefficient being approximately 40 times larger than the
value previously obtained in~\cite{Meissner:2023voo}, as can also be
seen from a comparison of Fig.~\ref{fig:fabund} with Fig.5 of
Ref.~\cite{Meissner:2023voo}.

If instead of the parameter sets ``NNLO'', ``fit'' and ``fit'' for
the reactions $p + \nuc{7}{Be} \to \nuc{8}{B} + \gamma$, $\nuc{3}{H} +
\nuc{4}{He} \to \nuc{7}{Li} + \gamma$ and $\nuc{3}{He} + \nuc{4}{He}
\to \nuc{7}{Be} + \gamma$, respectively, we use the parameter sets
``ANC'', ``A'' and ``AII'' of Tables~\ref{tab:reacpar}
and~\ref{tab:reacpar34} for these three reactions, respectively, we
find similar results, except for the $\nuc{7}{Li}\!\!+\!\!\nuc{}{Be}$ response
coefficients: In accordance with the fact that, as was shown in
Sect.~\ref{sec:alphadep}, the change of the rates with $\alpha$ was
found to be smaller for these parameters, the linear response
coefficient $c_1$ is about half as large and the quadratic coefficient
is smaller by a factor 2.5, still corresponding to an appreciable
curvature.

\section{\label{sec:summary}Summary} 

In this work we have studied the fine-structure constant dependence of
some BBN-relevant radiative capture reactions within the framework of
Halo-EFT.  We concentrated on the main effects, refraining from
implementing a coupled channel approach as would be dictated by strict
EFT power counting. Nevertheless we studied for each reaction two
parameter sets in order to obtain an indication of the systematic
errors. We found that the effects do deviate from what has been found
previously on the basis of parameterized cross section data and a
simple parameterization of the $\alpha$ dependence motivated by a
simple penetration factor. While for a neutron induced radiative
capture reaction the results are almost strictly linear, as is to be
expected since the radiative capture reaction amplitude is linear in
the electromagnetic coupling and thus the cross section is linear in
$\alpha$, for charged particle reactions the direct effect can both be
smaller, as is the case for the $\nuc{7}{Be}(p,\gamma)\nuc{8}{B}$
reaction, or larger, as is the case for the
$\nuc{4}{He}(\nuc{3}{H},\gamma)\nuc{7}{Li}$ and the
$\nuc{4}{He}(\nuc{3}{He},\gamma)\nuc{7}{Be}$ radiative captures, than
what is to be expected on the basis of the parameterized treatment.

In spite of these substantial deviations from the
$\alpha$-dependence of the parameterized rates obtained for these
reactions previously, the impact on the resulting abundances and on
their $\alpha$-dependence of the light elements $\nuc{2}{H}$,
$\nuc{3}{H}\!\!+\!\!\nuc{}{He}$, $\nuc{4}{He}$, $\nuc{6}{Li}$ with the rates
calculated within the framework of Halo-EFT is very minor only.  In
contrast for the $\nuc{7}{Li}\!\!+\!\!\nuc{}{Be}$-abundance we do find that
the $\alpha$-dependence differs appreciably from that of the previous
parameterized results, this $\alpha$-dependence being much more
pronounced and clearly non-linear with the Halo-EFT rates.  Also the
nominal abundance (\textit{i.e.} calculated with the current value of
the fine-structure constant $\alpha_0$) of $\nuc{7}{Li}\!\!+\!\!\nuc{}{Be}$ is
larger by almost 10 \%, whereas the other abundances remain
practically unchanged.

For reactions involving charged particles, the Halo-EFT calculation
accounts for the charged particle repulsion by inclusion of the full
Coulomb propagator in all reaction steps.  As the present study shows,
these Coulomb effects cannot always be approximated by a universal
penetration factor. It was also found that in some cases the study of
the fine-structure dependence of cross sections and the corresponding
rates within the framework of Halo-EFT can be limited by singularities
appearing in the normalisation, that enters as a factor in the
resulting cross sections. This was found to be relevant for the
$\nuc{3}{He} + \nuc{4}{He} \to \nuc{7}{Be} + \gamma$ reaction,
limiting the study to relative variations of $\alpha$ smaller than 
$6\%$\,. Furthermore, it should be stressed that the Halo-EFT framework
is of course restricted to those reactions where the di-nuclear
structure assumption underlying this is indeed applicable. Therefore
a definite assessment of the fine-structure dependence of rates relevant for
primordial nucleosynthesis should ultimately be performed within a framework
that allows for a genuine \textit{ab initio} treatment of nuclear reaction
dynamics. Indeed recent progress within the framework of nuclear lattice
effective field theory (NLEFT), see \textit{e.g.} Ref.~\cite{Elhatisari:2021eyg},
shows that NLEFT seems to be a promising candidate for such
a treatment.

\begin{acknowledgments}
  We thank Daniel Phillips for a useful communication.
  This project is part of the ERC Advanced Grant ``EXOTIC'' supported
  the European Research Council (ERC) under the European Union's Horizon
  2020 research and innovation programme (grant agreement
  No. 101018170), We further acknowledge support by the Deutsche
  Forschungsgemeinschaft (DFG, German Research Foundation) and the NSFC
  through the funds provided to the Sino-German Collaborative Research
  Center TRR110 ``Symmetries and the Emergence of Structure in QCD''
  (DFG Project ID 196253076 - TRR 110, NSFC Grant No. 12070131001), and
  the Chinese Academy of Sciences (CAS) President's International
  Fellowship Initiative (PIFI) (Grant No. 2018DM0034).
\end{acknowledgments}

\clearpage

\appendix

\section{\label{sec:CInt}$C$-integral}

To calculate
\begin{eqnarray}
  \label{eq:Cint}
  C(p)
  &=&
      \lim_{\delta\downarrow 0}\,
      \frac{\mu^2}{6\,\pi^2\,(p^2+\gamma^2)}
      \nonumber\\
  &&\times
      \intdif{0}{1}{x}
      \intdif{0}{1}{y}
      \frac{1}{\sqrt{x\,(1-x)}}
      \frac{1}{\sqrt{1-y}}
  \nonumber\\
  &&
     \Biggl(
     x\,p^2\,
     \log{\left[
     \frac{\pi}{4\,k_C^2}\left(
     -y\,p^2+(1-y)\frac{\gamma^2}{x}-\ii\,\delta
     \right)
     \right]}
  \nonumber\\
  &&
     +
     p^2\,
     \log{\left[
     \frac{\pi}{4\,k_C^2}\left(
     -y\,p^2-(1-y)\frac{p^2}{x}-\ii\,\delta
     \right)
     \right]}
  \nonumber\\
  &&
     +
     x\,\gamma^2\,
     \log{\left[
     \frac{\pi}{4\,k_C^2}\left(
     y\,\gamma^2+(1-y)\frac{\gamma^2}{x}-\ii\,\delta
     \right)
     \right]}
  \nonumber\\
  &&
     +
     \gamma^2\,
     \log{\left[
     \frac{\pi}{4\,k_C^2}\left(
     y\,\gamma^2-(1-y)\frac{p^2}{x}-\ii\,\delta
     \right)
     \right]}
     \Biggr)\,.
     \nonumber\\
\end{eqnarray}
Because the integral over $y$ can be performed analytically this
reduces to a single integral and, with the substitution $x=\sin\!^2{\vartheta}$\,, one obtains
with $\kappa = \gamma/p$ and $\eta_p = k_C/p$\,:
  \begin{equation}
    \label{eq:C0C1}
    \frac{C(p)}{\mu^2}
    =
    C_0
    +
    C_1(\kappa)\,,
  \end{equation}
  where
  \begin{equation}
    \label{eq:C0}
    C_0 = \frac{1}{2\pi}\,\log\left[\frac{\pi}{4\,\eta_p^2}\right]
  \end{equation}
  and
\begin{eqnarray}  
  \label{eq:C1}
  \lefteqn{
  C_1(\kappa)
  }
  \nonumber\\
  &=&
      \frac{1}{3\pi^2\left(1+\kappa^2\right)}
      \intdif{0}{\frac{\pi}{2}}{\vartheta}
      \Bigl\lbrace
      \sin\!^2{\vartheta}\,c_1(\kappa;\sin\!^2{\vartheta})
      +
      c_2(\kappa;\sin\!^2{\vartheta})
      \nonumber\\
  &&\hspace*{5em}
      +
      \sin\!^2{\vartheta}\,\kappa^2\,c_3(\kappa;\sin\!^2{\vartheta})
      +
      \kappa^2\,c_4(\kappa;\sin\!^2{\vartheta})
     \Bigr\rbrace\,,
     \nonumber
\end{eqnarray}
to be evaluated numerically with the integrands
\begin{eqnarray}
  &&\hspace*{-2em}
     c_1(\kappa;x)
     =
     \intdif{0}{1}{y}\,\,\frac{1}{\sqrt{1-y}}\,\log{\left[
     -y + (1-y)\,\frac{\kappa^2}{x}-\ii\,\delta
     \right]}
     \nonumber\\
  &=&
      2\,\log{\left[\frac{\kappa^2}{x}\right]}
      -
      4
      \nonumber\\
  &&\hspace*{1.5em}
     +
     \frac{2}{\sqrt{\frac{\kappa^2}{x}+1}}
     \left\lbrace
     \log{\left[
     \frac{\sqrt{\frac{\kappa^2}{x}+1}+1}{\sqrt{\frac{\kappa^2}{x}+1}-1}
        \right]}
     -\ii\,\pi
     \right\rbrace\,,
     \label{eq:c1}
\end{eqnarray}
\begin{eqnarray}
  &&\hspace*{-2em}
     c_2(\kappa;x)
     =
       \intdif{0}{1}{y}\,\,\frac{1}{\sqrt{1-y}}\,\log{\left[
     -y - (1-y)\,\frac{1}{x}-\ii\,\delta
     \right]}
     \nonumber\\
  &=&
      -2\,\log{\left[x\right]}
      -
      2\,\ii\,\pi
      -
      4
      \nonumber\\
  &&\hspace*{1.5em}
     +
     4\,\sqrt{\frac{x}{1-x}}\,
     \arctan{\left(\sqrt{\frac{1-x}{x}}\right)}\,,
     \label{eq:c2}
\end{eqnarray}
\begin{eqnarray}
  &&\hspace*{-2em}
     c_3(\kappa;x)
     =
     \intdif{0}{1}{y}\,\,\frac{1}{\sqrt{1-y}}\,\log{\left[
     \kappa^2\,y + (1-y)\,\frac{\kappa^2}{x}-\ii\,\delta
     \right]}
     \nonumber\\
  &=&
      2\,\log{\left[\frac{\kappa^2}{x}\right]}
      -
      4
      \nonumber\\
  &&\hspace*{1.5em}
     +
     4\,\sqrt{\frac{x}{1-x}}\,
     \arctan{\left(\sqrt{\frac{1-x}{x}}\right)}\,,
     \label{eq:c3}
\end{eqnarray}
\begin{eqnarray}
  &&\hspace*{-2em}
     c_4(\kappa;x)
     =
     \intdif{0}{1}{y}\,\,\frac{1}{\sqrt{1-y}}\,\log{\left[
     \kappa^2\,y - (1-y)\,\frac{1}{x}-\ii\,\delta
     \right]}
     \nonumber\\
  &=&
      -2\,\log{\left[x\right]}
      -
      2\,\ii\,\pi    
      -
      4
      \nonumber\\
  &&\hspace*{1em}
     +
     \frac{2\,\kappa}{\sqrt{\frac{1}{x}+\kappa^2}}
     \left\lbrace
     \log{\left[
     \frac{\sqrt{\frac{1}{x}+\kappa^2}+\kappa}{\sqrt{\frac{1}{x}+\kappa^2}-\kappa}
     \right]}
     +\ii\,\pi
     \right\rbrace\,.
     \label{eq:c4}
\end{eqnarray}

\section{\label{sec:DInt}$D'$-integral}

To calculate
\begin{eqnarray}
  \label{eq:Dint}
  D'(k_C;p,\gamma)
  &=&
      \lim_{\delta\downarrow 0}\,
      \frac{\mu^2}{6\,\pi^2\,(p^2+\gamma^2)}
      \nonumber\\
  &&\times
     \intdif{0}{1}{x}
     \intdif{0}{1}{y}
     \sqrt{\frac{x}{1-x}}
     \frac{1}{\sqrt{1-y}}
     \nonumber\\
  &&
     \Biggl(
     p^2\,
     \log{\left[
     \frac{1}{4\,k_C^2}\left(
     -y\,p^2+(1-y)\frac{\gamma^2}{x}-\ii\,\delta
     \right)
     \right]}
     \nonumber\\
  &&
     +
     p^2\,
     \log{\left[
     \frac{1}{4\,k_C^2}\left(
     -y\,p^2-(1-y)\frac{p^2}{x}-\ii\,\delta
     \right)
     \right]}
     \nonumber\\
  &&
     +
     \gamma^2\,
     \log{\left[
     \frac{1}{4\,k_C^2}\left(
     y\,\gamma^2+(1-y)\frac{\gamma^2}{x}-\ii\,\delta
     \right)
     \right]}
     \nonumber\\
  &&
     +
     \gamma^2\,
     \log{\left[
     \frac{1}{4\,k_C^2}\left(
     y\,\gamma^2-(1-y)\frac{p^2}{x}-\ii\,\delta
     \right)
     \right]}
     \Biggr)\,.
     \nonumber\\
\end{eqnarray}
Thus, with $\eta_p={k_C}/{p}$ and
$\kappa={\gamma}/{p}={\eta_p}/{\eta_\gamma}$\,:
\begin{eqnarray*}
  \lefteqn{
  \frac{D'(\eta_p,\kappa)}{\mu^2}
  =
  \lim_{\delta\downarrow 0}\,
  \frac{1}{6\,\pi^2\,(1+\kappa^2)}
  }
  \nonumber\\
&&\times
    \intdif{0}{1}{x}
    \intdif{0}{1}{y}
    \sqrt{\frac{x}{1-x}}
    \frac{1}{\sqrt{1-y}}
    \nonumber\\
&&
   \Biggl(
   \log{\left[
   \frac{1}{4\,\eta_p^2}\left(
   -y + (1-y)\frac{\kappa^2}{x}-\ii\,\delta
   \right)
   \right]}
   \nonumber\\
&&
   +
   \log{\left[
   \frac{1}{4\,\eta_p^2}\left(
   -y - (1-y)\frac{1}{x}-\ii\,\delta
   \right)
   \right]}
   \nonumber\\
&&
   +
   \kappa^2\,
   \log{\left[
   \frac{1}{4\,\eta_p^2}\left(
   y\,\kappa^2 + (1-y)\frac{\kappa^2}{x}-\ii\,\delta
   \right)
   \right]}
   \nonumber\\
&&
   +
   \kappa^2\,
   \log{\left[
   \frac{1}{4\,\eta_p^2}\left(
   y\,\kappa^2 - (1-y)\frac{1}{x}-\ii\,\delta
   \right)
   \right]}
   \Biggr)\,,
\end{eqnarray*}
\textit{i.e}
\begin{eqnarray}
  \label{eq:Dintp}
  \lefteqn{
  \frac{D'(\eta_p,\kappa)}{\mu^2}
  =
  -\frac{1}{3\,\pi}
  \,\log{\left(4\,\eta_p^2\right)}
  }
  \nonumber\\
&&
   +
   \frac{1}{3\,\pi^2\,(1+\kappa^2)}
   \intdif{0}{\frac{\pi}{2}}{\vartheta}
   \sin\!^2\vartheta
   \nonumber\\
&&\hspace*{2.0em}
   \Bigl\lbrace
   c_1(\kappa,\sin\!^2\vartheta)
   +
   c_2(\kappa,\sin\!^2\vartheta)
   \nonumber\\
&&\hspace*{3.0em}
   +
   \kappa^2\,c_3(\kappa,\sin\!^2\vartheta)
   +
   \kappa^2\,c_4(\kappa,\sin\!^2\vartheta)
   \Bigr\rbrace\,.
\end{eqnarray}
in terms of the integrands of Eqs.~\ref{eq:c1}-\ref{eq:c4} of Section~\ref{sec:CInt}\,.


\end{document}